\documentclass[useAMS,usenatbib,onecolumn,referee]{mnras}

\usepackage{dcolumn}
\usepackage{graphicx}
\usepackage{url}
\usepackage{color}
\usepackage{mathtools}
\usepackage{epstopdf}

\usepackage{longtable}

\usepackage{booktabs}

\newcommand{\cm}{cm$^{-1}$}

\newcommand{\ai}{\textit{ab initio}}
\newcommand{\Ai}{\textit{Ab initio}}

\newcommand{\ket}[1]{\vert #1 \rangle  }
\newcommand{\bra}[1]{\langle #1 \vert  }

\newcommand{\tnu}{\tilde{\nu}}

\newcommand{\trove}{{\sc TROVE}}

\title[ExoMol XXVII: Line list for Ethylene]{ExoMol molecular line lists  - XXVII:  spectra of C$_2$H$_4$}

\date{\today}

\author[B. P. Mant]{Barry P. Mant$^{1}$,  Andrey Yachmenev$^{2,3}$, Jonathan Tennyson$^{1}$\thanks{Email: j.tennyson@ucl.ac.uk}  and  Sergei N. Yurchenko$^{1}$ \\
$^{1}$ Department of Physics and Astronomy, University College London, London WC1E 6BT,
UK\\
$^{2}$ Center for Free-Electron Laser Science, Deutsches Elektronen-Synchrotron DESY, Notkestrasse 85,
22607 Hamburg, Germany\\
$^{3}$ The Hamburg Center for Ultrafast Imaging, Universitat Hamburg, Luruper Chaussee 149, ¨
22761 Hamburg, Germany}
\date{Accepted XXXX. Received XXXX; in original form XXXX}
\pubyear{2018}
\pagerange{\pageref{firstpage}--\pageref{lastpage}}
\begin{document}
\maketitle

\label{firstpage}

\begin{abstract}

  A new line list for ethylene, $^{12}$C$_2$$^1$H$_4$ is presented. The line list
  is based on high level \ai\ potential energy and dipole moment
  surfaces.  The potential energy surface is refined by fitting to
  experimental energies. The line list covers the range up to
  7000~\cm\ (1.43~$\mu$m) with all ro-vibrational transitions (50
  billion) with the lower state below 5000~\cm\ included and thus should be
  applicable for temperatures up to 700~K.  A technique for computing
  molecular opacities from vibrational band intensities is proposed
  and used to provide temperature dependent cross sections of ethylene
  for shorter wavelength and higher temperatures. When combined with
  realistic band profiles (such as the proposed three-band model), the
  vibrational intensity technique offers a cheap but reasonably
  accurate alternative to the full ro-vibrational calculations at high
  temperatures and should be reliable for representing molecular
  opacities. The C$_2$H$_4$ line list, which is called MaYTY, is
rmade available in
electronic form from the CDS
(\href{www.exomol.com}{http://cdsarc.u-strasbg.fr}) and ExoMol
(\href{www.exomol.com}{www.exomol.com}) databases.

\end{abstract}


\maketitle

\section{Introduction}

Hydrocarbons are an important class of molecules for planetary
atmospheres. Methane in particular has been detected in many places in
the solar system including: the atmospheres of Jupiter
\citep{96RaAlYu,03AtMaNi}, Saturn \citep{09GuFoBe}, Mars
\citep{07AtMaWo.CH4}, Uranus and Neptune \citep{93Lunine} as well as in
exoplanetary atmospheres \citep{08SwVaTi.exo,jt495}.
Methane is thought to be a key
biosignature \citep{93SaThCa.CH4}.  In atmospheres with an abundance
of methane, chemical reactions initiated by photolysis of C-H bonds
leads to the formation of larger hydrocarbons
\citep{96RaAlYu,09GuFoBe,14HuSexx}.  Particularly important are the
C$_2$H$_n$ hydrocarbons: acetylene, ethylene and ethane. These
molecules have been detected (along with propane, C$_3$H$_6$) in the
atmospheres of the solar system gas giants
\citep{96RaAlYu,03AtMaNi,09GuFoBe,93Lunine}. They have also been
observed in the atmosphere of Saturn's largest moon Titan
\citep{05NiAtBa} which has lakes of liquid hydrocarbons
\citep{07StElLu}. Hydrocarbons were even detected by the Cassini probe
in plumes from Enceladus \citep{06WaCoIp}.  Ethylene, the focus of
this work, is well-known in the in the circumstellar envelope of IRC+10216  \citep{81Betzxx.C2H4,17FoHiCe}
and is thought to be important in the atmospheres of exoplanets  \citep{13TiEnCo.exo}.

The ro-vibrational energy levels of ethylene have been the focus of multiple theoretical works in this decade.
This is due to both its importance and because it is one of the few 6-atom molecules which is
relatively rigid: the barrier to rotation of the CH$_2$ groups is 23~000 cm$^{-1}$ and involves breaking the
 $\pi$ bond \citep{98KrShBy}. This makes ethylene an ideal candidate to develop theoretical methods for medium sized molecules.
\citet{11AvCaxx.C2H4} calculated vibrational energies of C$_2$H$_4$ up to 4100 cm$^{-1}$ using
a basis pruning scheme and the Lanczos algorithm for obtaining the eigenvalues.
This was carried out using the quartic force field potential energy surface (PES) of \citet{95MaLeTa.C2H4}.
\citet{12CaShBo.C2H4} then built upon this work by calculating the ro-vibrational energies up to $J=40$ and
transition intensities using a dipole moment surface (DMS) computed at the MP2/aug-cc-pVTZ level of theory.
The ethylene molecule was also used as a test system to develop a new pruning approach by the same group \citep{15WaCaBo.C2H4}.
A new C$_2$H$_4$ PES \citep{14DeNiRe}  and DMS \citep{15DeNiRe} was recently constructed which gives
even more accurate energies and intensities. A high temperature line list was
subsequently constructed using these
surfaces by \citet{16ReDeNi}.


In this work we present new \textit{ab initio} potential energy and
dipole moment surfaces for ethylene and use them to compute a line
list for elevated temperatures as part of the ExoMol database project
\citep{jt528,jt631}. We name this line list MaYTY. Compared to the line list of \citet{16ReDeNi} we
slightly increase the applicable frequency range and include many more weak transitions (50 billion here compared to
60 million previously) which are important for total opacity.

Rovibrational energy
levels were computed variationally using a refined PES with the \trove\ program suite \citep{TROVE,15YaYuxx.method}.  Ethylene is the first
6 atom molecule in the Exomol database and the largest for which we have computed a
line list so far.

We also propose a new procedure for computing molecular opacities from vibrational transition moments only. Similar $J=0$ approaches are very common in simulating spectra of large polyatomic molecules \citep{12JoLaRo}, where either very simple band profiles (e.g. Lorentzian) or sophisticated functional forms (such as the narrow band approach of \citet{15CoLixx}) are used. Here we develop a three-band model based on three fundamental bands of C$_2$H$_4$ (one parallel and two perpendicular), which also represent its three dipole moment components. This $J=0$-effort approach has allowed us to significantly  extend the temperature as well as the frequency range of our line list and should be also useful for larger polyatomic molecules.

The paper is organised as follows: In Section \ref{sec:method} we give details of our PES and DMS along with
our variational calculations and how transition intensities were calculated. In Section \ref{sec:results} we give details
of the MaYTY line list and compare with experimental data. The new procedure for generating opacities from vibrational band intensities is discussed in Section \ref{sec:top-up}. We present conclusions in Sections \ref{sec:conclusions}.

\section{Methods}
\label{sec:method}

\subsection{Potential Energy Surface and Refinement}

An initial potential energy surface was constructed from \textit{ab initio} quantum chemistry calculations.
The explicitly correlated coupled cluster method CCSD(T)-F12b \citep{Adler07} was used with the F12-optimised correlation consistent polarized valence
cc-pVTZ-F12 basis set \citep{08PeAdWe} in the frozen core approximation. A Slater geminal exponent
of $\beta = 1.0$ $a_0^{-1}$ was used \citep{11HiPeKn}. For the resolution-of-the-identity
approximation to many-electron integrals we utilized the OptRI \citep{08YoKaPe.method} basis set, specifically matched to the cc-pVTZ-F12.
The additional many-electron integrals arising in the explicitly correlated methods
are calculated using the density fitting approach, for which we employed
cc-pV5Z/JKFIT \citep{Weigend02} and aug-cc-pwV5Z/MP2FIT \citep{Haettig05} auxiliary
basis sets.
All calculations were carried
out using MOLPRO2012 \citep{MOLPRO}.

\begin{figure}
\centering
\includegraphics[scale=0.3]{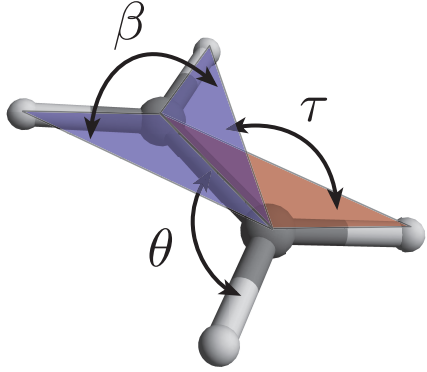}
\caption{Definition of angular internal coordinates in ethylene.}
\label{fig:coords}
\end{figure}

Electronic energies were calculated on a grid of 120\,000 molecular geometries for energies of up to $hc\cdot40\,000$~cm$^{-1}$
above the equilibrium geometry value. Up to eight of the twelve internal coordinates were varied at once.
The twelve coordinates used to represent the PES are: $\xi_1 = r_0 - r_0^{\rm eq}$ for the C--C bond stretching coordinate;
$\xi_j = r_i - r_1^{\rm eq}$ $j = 2,3,4,5$ for each of the C--H$_i$ ($i=1..4$) bond stretching coordinates;
$\xi_k = \theta_i - \theta_1^{\rm eq}$ $k = 6,7,8,9$, for each of the C--C--H$_i$ ($i=1..4$) valence angle bending coordinates;
$\xi_{10} = \pi - \beta_1$ and $\xi_{11} = \beta_2 - \pi$ where $\beta_{1}$ and $\beta_{2}$ are the two H--C--H book-type dihedral angles;
and $\xi_{12} = 2\tau - \beta_1 + \beta_2$ where $\tau$ is the dihedral angle between the two cis hydrogens.
For clarity, the angular internal coordinates are shown on Fig.~\ref{fig:coords}.
Values of $r_0^{\rm eq} = 1.331$ \AA, $r_1^{\rm eq} = 1.081$ \AA\ and $\theta_1^{\rm eq} = 121.45^{\circ}$ have been used.

The \textit{ab initio} energies were least squares fit to an analytical form consisting of long and short range parts as
\begin{equation}
\label{eq.pot_full}
f = V_{\text{short}} \times f_{\text{damp}} + V_{\text{long}}
\end{equation}
where $f_{\text{damp}}$ is a damping function to remove the contribution of the short range component at geometries where the
internal coordinates are far from their equilibrium values and has the form
\begin{align}
\label{eq.damp}
f_{\text{damp}} = & \exp \left(  -\delta_1 (\xi_1^2 + 2 \xi_1^4) \right) \times
                  \exp\left( -\delta_2 \sum_{i=2}^5 (\xi_i^2 + 2 \xi_i^4)  \right) \times \\
                  & \exp \left(-\delta_3 \sum_{i=6}^{9} (\xi_i^2 + 2 \xi_i^4) \right) \times  \exp \left(-\delta_4 \sum_{i=10}^{11} (\xi_i^2 + 2 \xi_i^4) \right) \times \\ & \exp \left(
                      -\delta_5 (\xi_{12}^2 + 2 \xi_{12}^4 ) \right).
\end{align}
The damping constants $\delta_i$ were kept fixed in the fitting at the values
$\delta_1=0.3$, $\delta_2=0.05$, $\delta_3=0.01$, $\delta_4=0.005$, $\delta_5=0.005$.
The long range function has the form
\begin{align}
  \label{eq.pot_long}
V_{\text{long}} = & \sum_{i=1}^5 D^{(2)}_i \xi_i'^2 + D^{(4)}_i \xi_i'^4 \\ \nonumber
                  & + \exp\left(-\gamma_1 \xi_1^2 - \gamma_2 \sum_{i=2}^5 \xi_i^2 \right)
                  \times  \sum_{i=6}^{12}\left( D_i^{(2)} \xi_i^2 +  D_i^{(4)} \xi_i^4 \right),
\end{align}
where $\gamma_1=0.1$ and $\gamma_2=0.1$ and the values for other parameters $D_i^{(2)}$ and $D_i^{(4)}$ were obtained in the least
squares fit to {\it ab initio} data points.
The short range function is a sum of symmetrised products of form
\begin{equation}
\label{eq.pot_short}
V_{\text{short}} = \sum_{i,j,k, \cdots} \left(\xi_1'^i \,\xi_2'^j \,\xi_3'^k \,\xi_4'^l \,\xi_5'^m \,\xi_6^n
\,\xi_7^o \,\xi_8^p \,\xi_9^q \,\xi_{10}^r \,\xi_{11}^s \,\xi_{12}^t \right)^\Gamma C_{ijk \cdots}
\end{equation}
where $\Gamma\equiv A_g$ produces symmetrized combinations of different permutations of the coordinates in the D$_{2h}$(M) molecular
symmetry group, $C_{ijk \cdots}$ are expansion parameters,
and Morse oscillator functions describe the stretching coordinates
$\xi_i' = 1 - \exp(-a_i \xi_i)$  with $a_1 = 1.88139$ \AA$^{-1}$ and $a_{2..5} = 1.79890$ \AA$^{-1}$.
The product was limited to a maximum of 8 coordinates coupled at the same time with the sum of powers
$i + j + \cdots + t \leq 8$. A total of 1269 terms were used in the sum.


The constants of the long range function in Eq.~(\ref{eq.pot_long}) and the expansion parameters of the short range potential
in Eq.~(\ref{eq.pot_short}) were found by least squares
fitting to the \textit{ab initio} energies.
Weight factors for energies were used as proposed by \cite{ps97}
\begin{equation}
  w_i = \left( \frac{{\rm tanh}[-0.0006\times (\tilde{E}_i-V_{\rm top})]+1.002002002}{2.002002002} \right)\times\frac{1}{N_i{\rm max}(\tilde{E}_i,V_{\rm lim})},
\end{equation}
where $hc \tilde{E}_i$ is the electronic energy at $i$-th geometry, $V_{\rm top}=30\,000$~cm$^{-1}$,
$V_{\rm lim}=15\,000$~cm$^{-1}$, and $N_i$ is the normalisation constant.
A weighted root-mean square (rms) error of 3.2 cm$^{-1}$
was obtained for energies up to $hc \cdot 40,000$ cm$^{-1}$. Expansion parameters and the explicit forms of the symmetrised
products in Eq.~\eqref{eq.pot_short} are given in a Fortran 90 subroutine in the supplementary information.

To improve the accuracy of nuclear motion calculations, the PES was refined using experimental data.
Refinement was carried out using a least-squares fitting procedure as implemented in \trove\ \citep{jt503} with the
pruned basis set (see below) in a very similar manner to that described in
a recent paper from our group \citep{jt701}.
Due to the large number of parameters used for the analytical representation of the PES and the size of the eigenfunctions for
ethylene, only parameters in Eq.~\eqref{eq.pot_short} with exponents summing to 2 were allowed to vary.
This includes linear ($\xi_i$),
harmonic ($\xi_i^2$) and mixed terms ($\xi_i \xi_j$) for a total of 21 parameters.
Refinement was carried out in two stages. First, 109 experimental
vibrational $J = 0$ band centres taken from \citet{99GeBaHe} were used.
This gave an initial refinement.
Then, 21 rotational-vibrational $J = 1$ energies from the HITRAN database \citep{jt691s}
were added and the refinement restarted.
Pure rotational energies were given the largest weights in the refinement of order 10$^4$ followed by $J = 1$
rotational-vibrational levels of order 10$^3$ and finally vibrational energies of order 0-10$^3$ depending on the reported
accuracy of these levels. Weights are normalised during the refinement and so only relative values are
important \citep{jt503}.

The refined PES was found to give accurate values for a further 155 $J = 2,3$ and 4 energy levels which were included, but
further iterations of refinement did not give improved values. This is due to both the size of the least-squares
fitting problem and that added rotational-vibrational levels were from the same vibrational bands as the $J = 1$ energies.
The difference between all observed energy levels used and the values given by our refined PES is shown 
in Fig.~\ref{fig:refplot}. The vibrational energies with observed$-$calculated errors of $>4$ cm$^{-1}$ were retained 
in the refinement to still provide some constraint to these states but were given relative weightings of a thousand times
 less than the HITRAN vibrational energies.

For the refined surface we obtained an rms error of 2.73 cm$^{-1}$ for the vibrational energies(reduced to
1.95 cm$^{-1}$ when bands which were given weights of zero in the refinement were excluded) compared to the values
quoted in \citet{99GeBaHe}. This is a large error but many of the bands included also gave
large errors for the global effective Hamiltonian model used by Georges \textit{et al.} and are of low accuracy. Bands
with the largest errors were given a weighing of zero in our fit. For the $J = 1$ data we obtain an rms error of
0.45 cm$^{-1}$ and 0.50 cm$^{-1}$ when all $J = 1-4$ is included respectively. When combined with the vibrational levels we
 obtain an overall rms of 1.75 cm$^{-1}$, which is reduced to
1.27 cm$^{-1}$ when bands which were given weights of zero in the refinement were excluded.

Table \ref{tab.vib_comp} compares fundamental vibrational band origins levels with empirical values for both our \ai\ and refined PES.
The energies were computed variationally using the basis set described in Section \ref{sec:variational}.

{\renewcommand{\arraystretch}{1.1}
\renewcommand{\tabcolsep}{0.5cm}
\begin{table}
\caption{Comparison of our computed vibrational fundamental band origins for C$_2$H$_4$ with the empirical
values of  \citet{99GeBaHe}. Energies in cm$^{-1}$.}
\begin{center}
\begin{tabular}{ll rrrcc}
\hline\hline
Band  &  Symmetry & \Ai\ PES& Refined PES & Obs.$^a$  & IR$^b$ & \trove$^c$ \\
\hline
      &           &        &                &          &        \\
$\nu_1$ &   $A_g$   &  3017.951 & 3021.797  & 3021.855 & & $n_2+n_3+n_4+n_5=1$   \\
$\nu_2$ &   $A_g$   &  1622.579 & 1625.648  & 1625.4   & & $n_1=1$   \\
$\nu_3$ &   $A_g$   &  1341.239 & 1342.361  & 1343.54  & & $n_6+n_7+n_8+n_9=1$  \\
$\nu_4$ &   $A_u$   &  1023.089 & 1024.610  & 1025.589 & & $n_{11}+n_{12}=1$   \\
$\nu_5$ &   $B_{3g}$&  3078.414 & 3084.426  & 3082.36  & & $n_2+n_3+n_4+n_5=1$    \\
$\nu_6$ &   $B_{3g}$&  1224.282 & 1227.050  & 1222     & & $n_6+n_7+n_8+n_9=1$   \\
$\nu_7$ &   $B_{3u}$&  947.846  & 948.830   & 948.770  & IR  & $n_6+n_7+n_8+n_9=1$  \\
$\nu_8$ &   $B_{2g}$&  937.202  & 939.069   & 939.86   & & $n_{11}+n_{12}=1$    \\
$\nu_9$ &   $B_{2u}$&  3100.809 & 3104.879  & 3104.872 & IR & $n_2+n_3+n_4+n_5=1$  \\
$\nu_{10}$& $B_{2u}$&  823.402  & 825.583   & 825.927  & IR & $n_6+n_7+n_8+n_9=1$  \\
$\nu_{11}$& $B_{1u}$& 2984.018  & 2988.709  & 2988.631 & IR & $n_2+n_3+n_4+n_5=1$   \\
$\nu_{12}$& $B_{1u}$& 1438.478  & 1441.725  & 1442.475 & IR  & $n_6+n_7+n_8+n_9=1$ \\
\hline\hline
\end{tabular}

\mbox{}\\
$^a$: \citet{99GeBaHe} \\
$^b$: Infrared active.  \\
$^c$: Correlation with the local mode (\trove) quantum numbers.

\label{tab.vib_comp}
\end{center}
\end{table}

\begin{figure}
\centering
\includegraphics[scale=0.6]{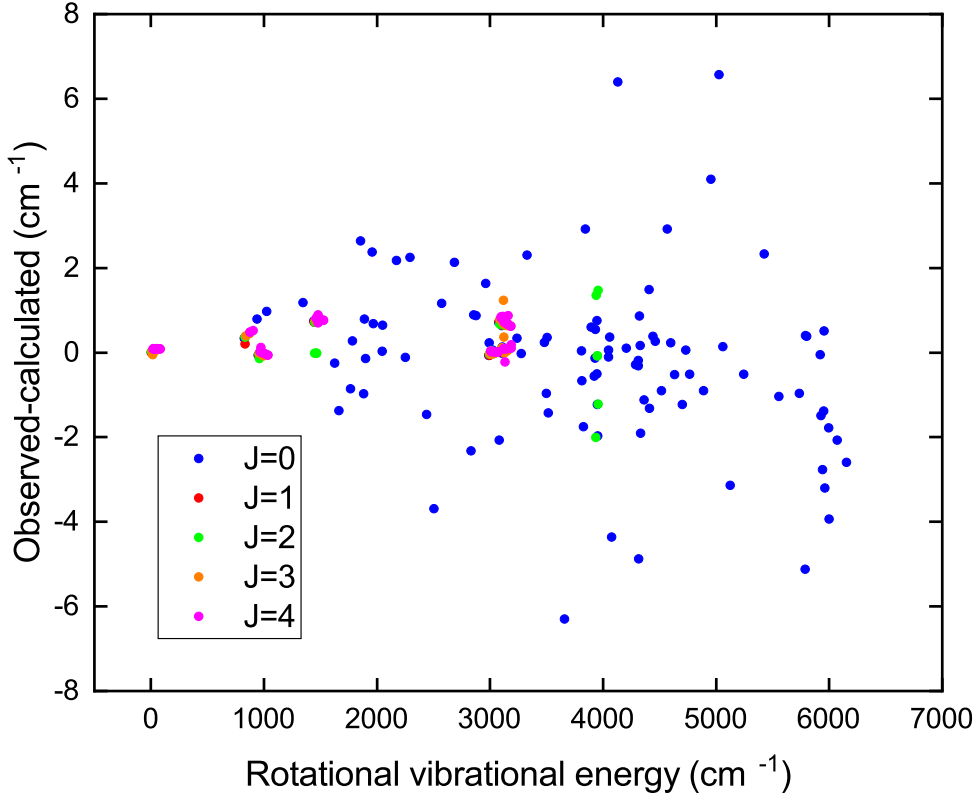}
\caption{Observed$-$Calculated rotational-vibrational energies after refinement. Vibrational energies (blue dots) with
larger discrepancies were given a 1000 times smaller weighting in the refinement with respect to HITRAN vibrational energies.}
\label{fig:refplot}
\end{figure}

As is usual for a refined PES used with \trove\ \citep{jt701,jt635,jt638,jt592,jt597,jt564}, the accuracy is only
guaranteed with the computational setup employed in this work.

\subsection{Dipole Moment Surface}

\textit{Ab initio} calculations for the DMS were carried out at the CCSD(T)-F12b/aug-cc-pVTZ
level of theory using the finite field method.
The frozen core approximation with a Slater geminal exponent $\beta=1.0$~a$_0^{-1}$
was employed using the same ansatz and auxiliary basis sets as the explicitly correlated
PES calculations.
For each of the $x$, $y$ and $z$ Cartesian
components an electric field of strength $\pm0.001$ a.u. was applied and the
dipole moment projections $\mu_x$, $\mu_y$ and $\mu_z$ computed as derivatives of the
electronic energy with respect to the field strength using central finite differences.
Calculations were carried out at about 93\,000 different molecular geometries
with energies up to $hc\cdot40\,000$~cm$^{-1}$, with up to six of the twelve
internal coordinates varied at once.

The DMS was fitted to an analytical form as follows. The origin of the molecule-fixed $xyz$ coordinate system ${\bf r}_O$
was taken to be the centre of the C$_1$--C$_2$ bond. The $z$-axis is chosen to be along the C$_1$--C$_2$ bond:
$$
\mathbf{e}_{z}  =  \frac{\mathbf{r}_{{\rm C}_1}-{\bf r}_O}{||\mathbf{r}_{{\rm C}_1}-{\bf r}_O||},
$$
where $\mathbf{r}_{{\rm C}_1}$ denotes Cartesian coordinates of carbon atom C$_1$.
 The $x$ axis is a symmetric combination average of the four normals to the four planes C$_1$C$_2$H$_i$ ($i=1,2,3,4$)
  as given by
 $$
\mathbf{e}_{x } =\frac{ \mathbf{e}_1 + \mathbf{e}_2 + \mathbf{e}_3 + \mathbf{e}_4 }{ ||\mathbf{e}_1 + \mathbf{e}_2 + \mathbf{e}_3 + \mathbf{e}_4||}
$$
and the $y$ axis is chosen as in the right-handed system. Here the normals are defined using the cross-products of the unit vector
$\mathbf{e}_{z}$ with the corresponding C--H bond vectors $\mathbf{e}_{{\rm H}_i}=(\mathbf{r}_{{\rm H}_i}-{\bf r}_O)/||\mathbf{r}_{{\rm H}_i}-{\bf r}_O||$ ($i=1..4$)
 as given by
\begin{align}
\label{eq.}
 \nonumber
\mathbf{e}_1 =  -\frac{ \mathbf{e}_{z} \times \mathbf{e}_{{\rm H}_1} }{|| \mathbf{e}_{z} \times \mathbf{e}_{{\rm H}_1} ||}, & \quad
\mathbf{e}_2 =  \frac{ \mathbf{e}_{z} \times \mathbf{e}_{{\rm H}_2} }{|| \mathbf{e}_{z} \times \mathbf{e}_{{\rm H}_2} ||}, \\
\nonumber
\mathbf{e}_3 =  -\frac{ \mathbf{e}_{z} \times \mathbf{e}_{{\rm H}_3} }{|| \mathbf{e}_{z} \times \mathbf{e}_{{\rm H}_3} ||}, & \quad
\mathbf{e}_4 =  \frac{ \mathbf{e}_{z} \times \mathbf{e}_{{\rm H}_4} }{|| \mathbf{e}_{z} \times \mathbf{e}_{{\rm H}_4} ||}    ,
\end{align}
where $\mathbf{r}_{{\rm H}_i}$ denotes Cartesian coordinates of hydrogen atoms.
The Cartesian axes $x$, $y$ and $z$ transform according to D$_{2h}$(M) as $B_{3u}$,  $B_{2u}$ and $B_{1u}$ irreducible representations (irreps), respectively.



The dipole moment vector can be expressed as
\begin{equation}
\label{eq.dipolevec}
{\boldsymbol\mu} = \mu_x \mathbf{e}_x + \mu_y \mathbf{e}_y + \mu_z \mathbf{e}_z,
\end{equation}
where $\mu_{\alpha}$ ($\alpha = x,y,z$) are functions of the internal coordinates of the form
\begin{equation}
\label{eq.dipole_func}
 \mu_{\alpha} = \sum_{i,j,k,\cdots} \left(\xi_1^i \, \xi_2^j \, \xi_3^k \, \cdots \xi_{12}^t \right)^\Gamma F^{(\alpha)}_{ijk \cdots}
\end{equation}
where $\Gamma$ produces symmetrized combinations of different permutations of the coordinates in the $B_{3u}$, $B_{2u}$ and
$B_{1u}$ irreps for $\alpha=x$, $y$ and $z$, respectively, and $F^{(\alpha)}_{ijk \cdots}$ are the expansion parameters.
The symmetry-adapted analytical expressions in Eq.~\eqref{eq.dipole_func}
have been obtained using the {\sc SymPy} Python library for symbolic mathematics \citep{SymPy}.
The Python program is freely available from the authors upon request.
The coordinates $\xi_1,\xi_2,...\xi_{12}$ chosen for analytical representation of the DMS in Eq.~\eqref{eq.dipole_func}
are the same as those used for the PES.

For each component of the dipole we used  a sixth-order expansion consisting of 1881, 1861 and 1399 terms for the $B_{1u}$,
$B_{2u}$ and $B_{3u}$ symmetries, respectively. The expansion parameters $F^{(\alpha)}_{ijk \cdots}$ were determined by least squares
fitting to the \textit{ab initio} data giving rms errors of $7\times10^{-4}$, $7\times10^{-4}$ and $6\times10^{-4}$ D,
respectively.
The expansion parameters $F^{(\alpha)}_{ijk \cdots}$ and Fortran 90 subroutines  to compute $\mu_\alpha$ are provided as part of the supplementary information.

\subsection{Variational Calculations}
\label{sec:variational}

Variational ro-vibrational calculations were carried out using the \trove\ program. The \trove\ methodology is well documented
 \citep{TROVE,jt466,15YaYuxx.method,17YuYaOv.methods,jt626} and has been applied to a variety of molecules as part of the
ExoMol project
 \citep{jt466,jt500,jt556,jt553,jt554,jt564,jt592,
jt597,jt620,jt612,jt641,jt701}.
Only the specific details used in this work on ethylene will be discussed here.

The ro-vibrational Hamiltonian was constructed numerically via an automatic differentiation method \citep{15YaYuxx.method}.
The Hamiltonian was expanded using a power series in curvilinear coordinates around the  equilibrium geometry of the molecule.
The coordinates used were the same as those used to fit the PES.


The kinetic energy operator was expanded to the 6th order and the potential energy operator to the 8th order.
The same Morse coordinates as used in Eq.~\eqref{eq.pot_short} were used for the potential expansion for the
stretching coordinates ($i = 1-5$) with the other bending coordinates expanded as $\xi_i$ themselves.
Atomic masses were used throughout.

A multistep contraction scheme was used to build the vibrational basis set. For each coordinate a one-dimensional
Schr\"{o}dinger equation was solved using the Numerov-Cooley approach~\citep{24Nuxxxx.method,61Coxxxx.method,TROVE} to
generate basis functions $\phi_{n_i} (\xi_i)$ with vibrational quantum number $n_i$.
The vibrational basis set functions $\ket{v}$ are
formed as products of the 1D basis functions
\begin{equation}
\label{eq.vibprod}
\ket{v} = \prod_{\nu} \ket{n_{\nu}} = \phi_{n_1} (\xi_1)\phi_{n_2}(\xi_2)\ldots \phi_{n_{12}} (\xi_{12}) .
\end{equation}
The basis set is truncated by the polyad number $P$ via
\begin{equation}
\label{eq:polymax}
P = n_1 + 2(n_2 + n_3 + n_4 + n_5) + n_6 + n_7 + n_8 + n_9 + n_{10} + n_{11} + n_{12} \leq P_{\text{max}}.
\end{equation}
A value of $P_{\text{max}} = 10$ was used. This is a smaller value than used for previous Exomol line lists \citep{jt635,jt592,jt701,jt564,jt638,jt597}
but with 12 degrees-of-freedom the basis set rapidly increases with increasing $P$. To estimate the converge of
vibrational energies with this basis set we used a complete vibrational basis set (CVBS) extrapolation procedure similar to
that described by \citet{jt612}. Variational calculations were carried out with $P_{\text{max}} = 6, 8$ and 10
respectively. From this we estimate that above 4000 cm$^{-1}$ there are some vibrational levels (typically those with multiple
bending modes excited) which are only converged to around 4 cm$^{-1}$ with a $P_{\text{max}} = 10$ basis. The average
convergence error for 0--5~000 cm$^{-1}$ is estimated to be only 1.5 cm$^{-1}$ however. It should be noted that these estimates do not account for the fact that the  PES refinement procedure described above tends to compensate partly or fully for the basis set convergence errors, even when extrapolating to higher vibrational excitations. Strictly speaking, in order to get a sensible convergence error, one would need to produce a refined PES for all three values of $P_{\text{max}} = 6, 8$ and 10. These estimates do however indicate the possible error of our effective ($P=10$) PES if used with larger basis sets or other
nuclear motion methods.

 To extend the vibrational basis, further
1D and 2D functions were added with $\phi_{n_i} (\xi_i)$ ($n = 11,12,13,14$) and $\phi_{n_i} (\xi_i) \phi_{n_j} (\xi_j)$
($ 10 \leq i + j  \leq 14$). A contracted basis set was then formed by reducing the 12 dimensional problem into
5 subspaces: ($\xi_1$), ($\xi_2, \xi_3, \xi_4, \xi_5$), ($\xi_6, \xi_7, \xi_8, \xi_9$), ($\xi_{10}, \xi_{11}$)
and ($\xi_{12}$). A Hamiltonian matrix is constructed and diagonalised for each of these subspaces to
give symmetrised contracted vibrational basis functions. The details of this step have been discussed in a
recent publication \citep{17YuYaOv.methods}. Products of these eigenfunctions are formed which are also truncated via
 Eq.~\eqref{eq:polymax}.

 To increase the computational efficiency of this step, a new algorithm
 for sorting and calculating matrix elements of the PES between
 primitive basis functions was implemented. This procedure also sets these elements to zero for potential expansion coefficients with  values smaller than a tolerance factor. Here we take this as 0.01 (in the units of \cm, Angstrom and radian).  This procedure led to around
 a 70 fold speed up for smaller basis test calculations whilst only
 affecting the accuracy of vibrational states by 0.01 cm$^{-1}$, far
 lower than the error of the \textit{ab initio} PES.  This new
 `fast-ci' method will be described fully in a subsequent publication.

Following this procedure, 145~240 vibrational  eigenfunctions $\ket{\Phi_{\text{vib}}^{(i)} }$ of C$_2$H$_4$ were obtained with term energies $\tilde{E}_{i}^{(J=0)}$ up to 21~000 cm$^{-1}$ above the ground state (our post refinement zero-point-energy is
 11~022.5~\cm). According to the $J=0$-contraction scheme \trove\ uses these $J=0$ eigenfunctions as the vibrational basis set.
However using a basis set of this size for high rotationally excited
levels is currently impractical and it was necessary to reduce the number of basis functions. The basis set was further truncated using the same
approach  based on the vibrational band intensity as described in a recent paper for the silane (SiH$_4$) line list \citep{jt701}, which will be referenced to as intensity basis set pruning  (IBSP).
According to this approach, the vibrational basis functions $\ket{\Phi_{\text{vib}}^{(i)} }$ above some energy threshold, $\tilde{E}_{\rm max}^{(J=0)}$, should be truncated with the exception of  functions responsible for significant contribution to the absorption opacity (larger than some intensity threshold $I_{\rm max}$). In turn, the absorption contribution is estimated from the intensities of the corresponding bands using  these functions as the upper or lower states.

We define the vibrational absorption intensity (cm$/$molecule) for the band $f\gets i$ as
\begin{equation}
\label{eq.vibint}
I_{fi}^{(J=0)} = \frac{A_{fi}^{(J=0)}}{8 \pi c}  \frac{\exp(- c_2 \tilde{E}_i^{(J=0)}/T) }{Q^{(J=0)} \left(\tnu^{(J=0)}_{fi}\right)^2}.
\end{equation}
The vibrational Einstein coefficient (s$^{-1}$) is given by
\begin{equation}
\label{eq.vib-einsteinA}
A_{fi}^{(J=0)} = \frac{8\times 10^{-36} \pi^4 \left(\nu^{(J=0)}_{fi}\right)^3}{3h} \bar\mu_{fi}^2,
\end{equation}
where the vibrational transition moment $\mu_{fi}$ (D) is
\begin{equation}
\label{eq.vibtm}
 \bar\mu_{fi}  = \sqrt{ \sum_{\alpha = x, y, z} \left|
 \bra{\Phi_{\text{vib}}^f } \bar{\mu}_{\alpha} \ket{\Phi_{\text{vib}}^i } \right| ^2  }.
\end{equation}
Here $h$ is Planck's constant,  $Q^{(J=0)}$ is the vibrational ($J=0$) partition function,
$\tilde{E}_i^{(J=0)}$ and $\tilde{\nu}_{fi}^{(J=0)}$ are the
vibrational  lower state term value and band centres,
respectively and $c_2$ is the second radiation constant.  Here
$\ket{\Phi_{\text{vib}}^{(i)} }$ and $\bra{\Phi_{\text{vib}}^{(f)} }$
are the initial and final state vibrational eigenfunctions,
respectively and $\bar{\mu}_{\alpha}$ is the electronically averaged
dipole moment along the molecular fixed axis $\alpha = x,y,z$.


The vibrational absorption intensities were
computed between each state at an elevated temperature of 800~K. For
each vibrational state, the largest intensity to or from that state
was then associated with that state. The $J = 0$ basis set was then
pruned based on this. All states up to $hc \cdot$8000 cm$^{-1}$ were
retained. States with energy above this were discarded if their
largest intensity was less than some value $I_{\text{max}}$. Here a
value of $I_{\text{max}} = 1 \times 10^{-24}$ cm$/$molecule was used.
This value was chosen to retain as many states as possible (which
support intense transitions) whilst making the calculations for high
$J$ practical.  The resulting pruned vibrational basis contained 13~572 functions
corresponding to energies up to $hc \cdot 12~000 $ cm$^{-1}$.  This
basis was then used for $J > 0$ calculations by combining it with
symmetrized rigid-rotor functions as described previously
\citep{jt466,17YuYaOv.methods}.

The pruning procedure based on the $J$=$0$-contraction has the advantages that the accuracy of the vibrational energy levels and eigenfunctions computed
using the unpruned basis is retained. The errors introduced in pruning the basis for the ro-vibrational levels are compensated for by refining the PES with the pruned basis.

Fig. \ref{fig:vibcross} shows vibrational intensities of C$_2$H$_4$
computed using Eq.~\eqref{eq.vibint} for $T=500$~K as cross sections.
Here we compare the total cross sections (no pruning) and the
contribution missing due to the intensity-based pruning.  The effect
of the pruning on the intensities is negligible for the range below
7000~\cm\ ($\sim$~0.01~\%). This is especially important
for hot spectra applications, where the completeness of the molecular
absorption arguably plays a more important role than the accuracy \citep{jt572}.

\begin{figure}
\centering
\includegraphics[scale=0.5]{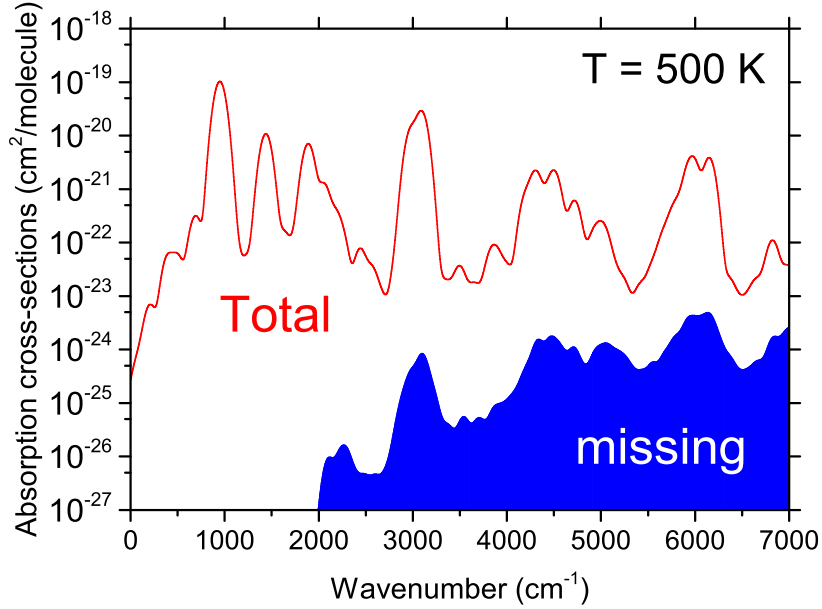}
\caption{The vibrational cross sections: total compared to the missing contribution due to the vibrational pruning.}
\label{fig:vibcross}
\end{figure}

A final empirical basis set correction was also made to shift vibrational band centres to experimental values. As the
$J=0$ basis is diagonal with respect to the vibrational component of the Hamiltonian, experimental band centres can be
used instead \citep{jt503}. This was carried out for bands with observable Q branches namely: the $\nu_{12}$ band at
1442.4750 cm$^{-1}$ \citep{98RuFiKh.C2H4}, the $\nu_2 + \nu_{12}$ band at 3078.46270 cm$^{-1}$ \citep{98RuFiKh.C2H4},
$\nu_7 + \nu_{8}$ band at 1888.9748 cm$^{-1}$ \citep{98RuFiKh.C2H4},
 $\nu_6 + \nu_{10}$ band at 2047.76 cm$^{-1}$ \citep{98RuFiKh.C2H4}
 and $\nu_5 + \nu_9$ band at 6150.98104 cm$^{-1}$ \citep{99BaGeHe}.

For ro-vibrational states with $0 \le J \le 24$, variational calculations were carried out using \trove\ in a standard manner.
That is, the Hamiltonian was constructed and diagonalised with all data kept in RAM memory.
The largest matrix to be diagonalised in this case had around 82~000 rows per symmetry for $J=24$. For states with $J>24$
 however, this was not possible. Instead the procedure described by \citet{jt641} for the SO$_3$ line list was used. This involved first calculating the Hamiltonian and then saving to disk. Diagonalisation was then
carried out using an MPI-optimized version of the eigensolver, PDSYEVD. This was carried out separately for each $J$ and symmetry
$\Gamma$, where $\Gamma$= $A_{1g},A_{1u},B_{1g},B_{1u},B_{2g},B_{2u},B_{3g}$ and $B_{3u}$. A final run of \trove\ is required to reformat the eigenvalues and eigenvectors into the proper format. The
largest matrix diagonalised had 150~000 rows for $J = 50$. For $J>50$ the Hamiltonian decreases in size as we only retain
eigenvalues less than 18~000 cm$^{-1}$.

\subsection{Line Intensities}

The eigenvectors from the variational calculation along with the DMS were used to compute Einstein-A coefficients of
transitions. These satisfy the rotational selection rules \citep{98BuJexx}
\begin{equation}
\label{eq.rotselect}
J' - J'' = 0, \pm 1, \quad {\rm and} \quad J' + J'' \neq 0,
\end{equation}
where $J'$ and $J''$ are the upper and lower values of the total angular quantum number $J$ and symmetry selection rules
\begin{equation}
\label{eq.symselect}
A_{1g} \leftrightarrow A_{1u}, B_{1g} \leftrightarrow B_{1u}, B_{2g} \leftrightarrow B_{2u},
B_{3g} \leftrightarrow B_{3u}.
\end{equation}

The absolute absorption intensities are then given by \citep{98BuJexx}
\begin{equation}
\label{eq.intensity}
I(f \leftarrow i) = \frac{A_{fi}}{8 \pi c} g_{ns} (2 J_f + 1) \frac{\exp(- c_2 \tilde{E}_i/T) }{Q(T) \tnu^2_{fi}}
\times \left[ 1 - \exp\left( - \frac{c_2 \tnu_{fi}}{T} \right) \right],
\end{equation}
where  $J_f$ is the rotation quantum number for the final state, $\tnu_{fi}$ is the
transition frequency ($\tnu_{fi} = \tilde{E}_f - \tilde{E}_i$), $\tilde{E}_i$ and $\tilde{E}_f$ are the initial and upper state term values, respectively, and
$Q(T)$ is the partition function (Section \ref{sec:part}). The Einstein-A coefficients $A_{fi}$  between the ro-vibrational states $i$ and  $f$ are defined in \citet{05YuThCa.method}.
The nuclear spin statistical weights $g_{ns}$ for ethylene are (7,7,3,3,3,3,3,3) for states of symmetry
($A_{1g},A_{1u},B_{1g},B_{1u},B_{2g},B_{2u},B_{3g},B_{3u}$) respectively \citep{98BuJexx} within
the HITRAN convention, adopted by ExoMol, of including the full nuclear spin of each species.

The temperature independent Einstein-A coefficients were computed using the GAIN-MPI program \citep{jt653}.
With this program we were able to calculate up to 93~000 transitions per second using NVIDIA Tesla P100 GPUs on the Wilkes2
cluster.

Intensities were computed using a lower energy range of 0 -- 5000 cm$^{-1}$ taking into account up to $J = 78$ for transitions
frequencies between 0 and 7000 cm$^{-1}$. An intensity cut-off of 10$^{-50}$ cm molecule$^{-1}$ at $T=298$~K was used,
ensuring that essentially all transitions are taken into account for up to around 700 K (see section \ref{sec:part}).

\section{Results}
\label{sec:results}

\subsection{Partition Function}
\label{sec:part}

The temperature-dependent partition function $Q(T)$ is defined as
\begin{equation}
\label{eq.part_func}
Q(T) = \sum_i  g_i \exp\left( - c_2\frac{ \tilde{E}_i}{T} \right),
\end{equation}
where $g_i = g_{ns}(2J_i + 1)$ is the degeneracy of the state $i$ with energy $E_i$ and
rotational quantum number $J_i$.

Fig. \ref{fig:compare_part} shows the convergence of $Q(T)$ as a function of $J$ at different
temperatures. At 700 K the partition function is converged to 0.02\%.
In Table \ref{tab.part_func} we compare the partition function calculated at various temperature with those of
literature values. In general agreement between the various sources is good. Our value which increases
slightly faster with temperature than those of
\citet{16ReDeNi} is probably due to our more complete treatment of the energy levels. In the supplementary information we
provide the partition function between 0 and 1500 K at 1 K intervals.

The current line list was computed with a lower energy threshold of
$hc \cdot 5000$ \cm. To assess the completeness of our line list we
compute a reduced partition function, $Q_{\text{limit}}$ which only
takes into account energies up to $hc \cdot5000$ \cm\ in
Eq.~\eqref{eq.part_func}. Fig. \ref{fig:compare_part} also shows a
plot of the ratio of $Q_{\text{limit}}/Q$. At 700~K the ratio is 0.98
and this temperature can be taken as a soft limit. At higher
temperatures opacity will progressively be underestimated (see Section
\ref{sec:top-up}).

{\renewcommand{\arraystretch}{1.1}
\renewcommand{\tabcolsep}{0.5cm}
\begin{table}
\caption{Comparison of partition functions with literature values.}
\begin{center}
\begin{tabular}{c c c c c l}
\hline
$T$(K) &  Refs & $Q_{\text{vib}}$ & $Q_{\text{rot}}$ & $Q_{\text{vib}}Q_{\text{rot}}$ & Direct Sum \\
\hline
     &             &      &      &     &             \\
80   & This work   &    1  &  1472.7    &  1472.7   &     1479.7  \\
     & \citet{01BlHiFa.C2H4} & 1 &  1467.9  & 1467.9  &  \\
     & \citet{jt692}      &      &      &  1474.2   &        \\
     & \citet{16ReDeNi} & 1 & 1471.7 & 1471.7 & 1471.7 \\
     &             &      &      &            &   \\
160  & This work   &   1.0011   &  4169.6    &  4174.4   &    4181.4    \\
     & \citet{01BlHiFa.C2H4} & 1.0009 &  4143.3  & 4147.1   &  \\
     & \citet{jt692}      &      &      &  4169.4   &        \\
     & \citet{16ReDeNi} & 1.0011 & 4154.0 & 4158.6 & 4158.8 \\
          &             &      &      &            &   \\
296   & This work   &   1.0522   &  10498.3    &  11046.3   &   11058.6    \\
     & \citet{01BlHiFa.C2H4} & 1.0469 & 10421.4  & 10910.1  &  \\
     & \citet{jt692}      &      &      &  11041.5   &        \\
     & \citet{12CaShBo.C2H4} &   &   &   &  10979.2 ($J<40$) \\
     & \citet{08RoBoAu.C2H4} & 1.0469 & 10432.9 & 10922.2 &  \\
     & \citet{16ReDeNi} & 1.0521 & 10448.4 & 10992.8 & 10997.8 \\
          &             &      &      &            &   \\
500    & This work    & 1.4402  &  23069.5  &  33224.8  & 33306.1   \\
       & \citet{jt692} &  &  & 33271.3 &  \\
     &  \citet{16ReDeNi} & 1.4396  & 22951.3  &  33040.7 & 33117.2 \\
          &             &      &      &            &   \\
700  & This work  & 2.4298 &  38251.5  & 92942.0  &  93373.4  \\
     & \citet{jt692}     &  &   & 93244.0  &  \\
     &  \citet{16ReDeNi} & 2.4274 & 38048.0 & 92357.7 & 92702.8 \\
\hline
\end{tabular}
\label{tab.part_func}
\end{center}
\end{table}

\begin{figure}
\centering
  \begin{tabular}{cc}
    \includegraphics[width=.5\textwidth]{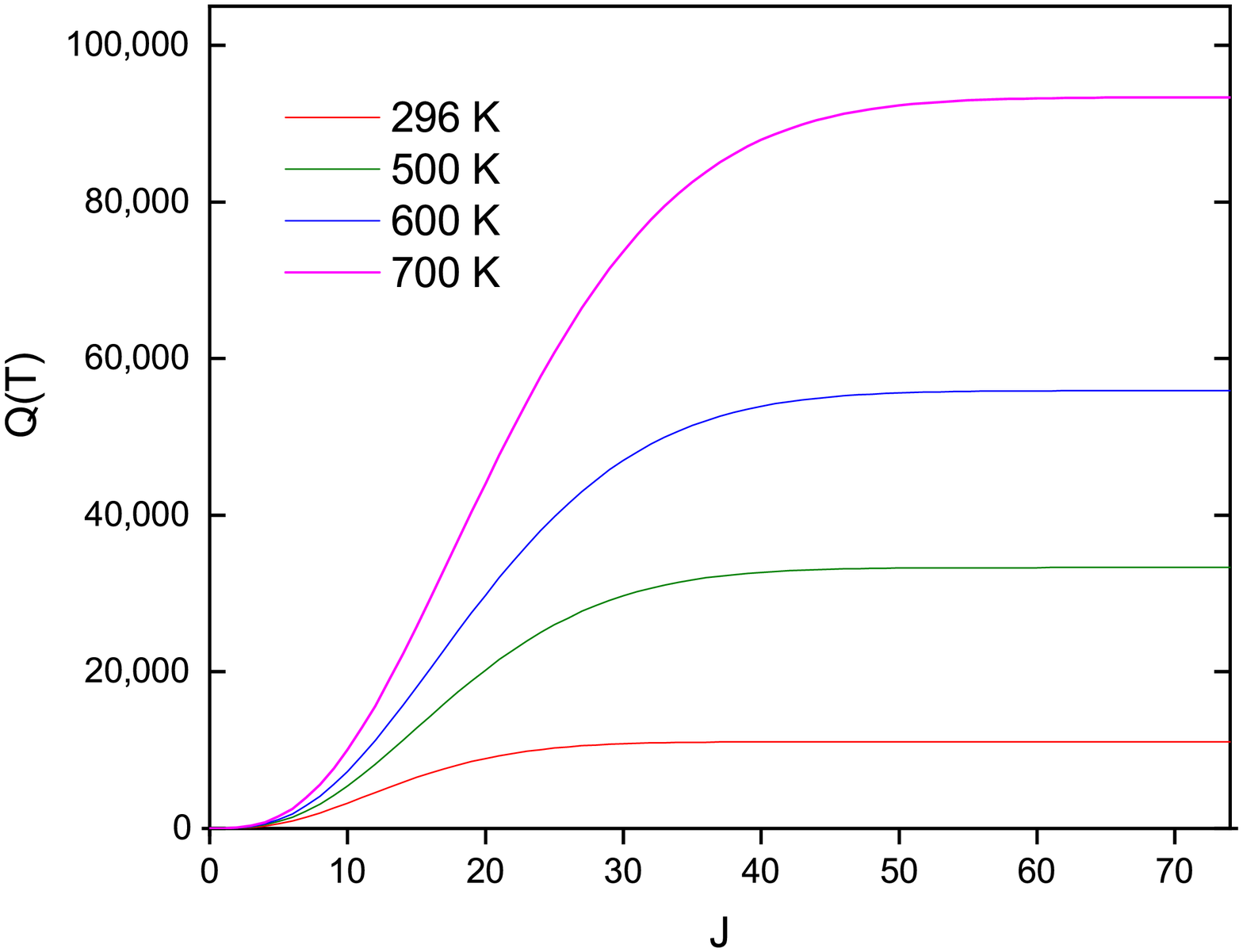} &
    \includegraphics[width=.5\textwidth]{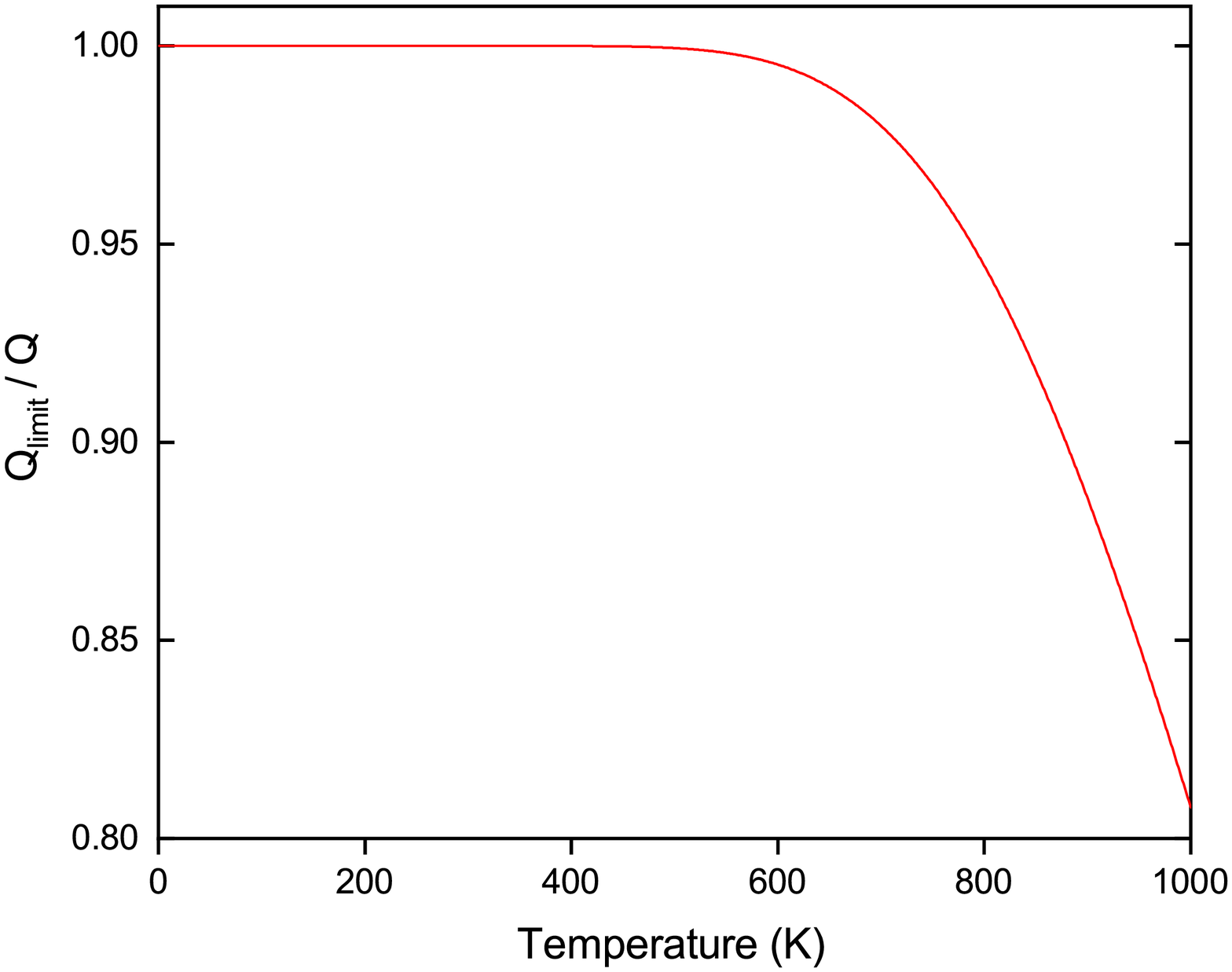} \\
  \end{tabular}
  \caption{Partition functions of C$_2$H$_4$: Convergence of partition function $Q(T)$ at different temperatures with respect
   to the rotational quantum number $J$ (left) and $Q_{\text{limit}}/Q$ as a function of temperature (right).}
  \label{fig:compare_part}
\end{figure}

\subsection{Line List Format}

A complete description of the ExoMol data structure along with examples was reported by \citet{jt631}.
The .states file contains all computed ro-vibrational energies (in cm$^{-1}$) relative to the ground state. Each energy level
is assigned a unique state ID with symmetry and quantum number labelling as shown in Table \ref{tab.tran}. The .trans files,
which are split into frequency windows for ease of use, contain all computed transitions with upper
and lower state ID labels, and Einstein A coefficients. An example from a .trans file for the line list is given in
Table \ref{tab.tran}.

{\renewcommand{\arraystretch}{1.12}
\renewcommand{\tabcolsep}{0.15cm}
\begin{table}
\caption{Extract from  .states file for MaYTY line list.}
\begin{center}
\begin{tabular}{c r c c c c c c c c c c c c c c c c c c c c}
\hline
$N$ & \multicolumn{1}{c}{$\tilde{E}$} & $g_{\text{tot}}$ &  $J$ & $\Gamma_{\text{tot}}$ & $n_1$ & $n_2$ & $n_3$ & $n_4$ & $n_5$ & $n_6$ & $n_7$ & $n_8$ & $n_9$ & $n_{10}$ & $n_{11}$ & $n_{12}$ & $\Gamma_{\text{vib}}$ & $J$ & $K$ & $\tau_{\text{rot}}$ & $\Gamma_{\text{rot}}$  \\
\hline
 &              &   &   &   &   &   &   &   &   &   &   &   &   &   &   &   &   &   &   &   &   \\
1 & 0.000000    & 7 & 0 & 1 & 0 & 0 & 0 & 0 & 0 & 0 & 0 & 0 & 0 & 0 & 0 & 0 & 1 & 0 & 0 & 0 & 1 \\
2 & 1342.361058 & 7 & 0 & 1 & 0 & 0 & 0 & 0 & 0 & 1 & 0 & 0 & 0 & 0 & 0 & 0 & 1 & 0 & 0 & 0 & 1 \\
3 & 1625.647919 & 7 & 0 & 1 & 1 & 0 & 0 & 0 & 0 & 0 & 0 & 0 & 0 & 0 & 0 & 0 & 1 & 0 & 0 & 0 & 1 \\
4 & 1663.565812 & 7 & 0 & 1 & 0 & 0 & 0 & 0 & 0 & 1 & 0 & 1 & 0 & 0 & 0 & 0 & 1 & 0 & 0 & 0 & 1 \\
5 & 1881.875767 & 7 & 0 & 1 & 0 & 0 & 0 & 0 & 0 & 0 & 0 & 0 & 0 & 0 & 2 & 0 & 1 & 0 & 0 & 0 & 1 \\
6 & 1899.881666 & 7 & 0 & 1 & 0 & 0 & 0 & 0 & 0 & 0 & 0 & 0 & 0 & 1 & 1 & 0 & 1 & 0 & 0 & 0 & 1 \\
7 & 2046.368511 & 7 & 0 & 1 & 0 & 0 & 0 & 0 & 0 & 0 & 0 & 0 & 0 & 0 & 0 & 2 & 1 & 0 & 0 & 0 & 1 \\
8 & 2452.971198 & 7 & 0 & 1 & 0 & 0 & 0 & 0 & 0 & 1 & 0 & 0 & 1 & 0 & 0 & 0 & 1 & 0 & 0 & 0 & 1 \\
9 & 2683.172133 & 7 & 0 & 1 & 1 & 0 & 0 & 0 & 0 & 1 & 0 & 0 & 0 & 0 & 0 & 0 & 1 & 0 & 0 & 0 & 1 \\
10& 2784.166877 & 7 & 0 & 1 & 0 & 0 & 0 & 0 & 0 & 0 & 0 & 1 & 0 & 0 & 1 & 1 & 1 & 0 & 0 & 0 & 1 \\
\hline
\end{tabular}
\label{tab.states}
\end{center}
$N$ : State ID;

$\tilde{E}$: Term value (in cm$^{-1}$);

$g_{\text{tot}}$: Total degeneracy;

$J$: Total angular momentum;

$\Gamma_{\text{tot}}$: Total symmetry in D$_{2h}$(M) (1 is $A_g$, 2 is $A_u$, 3 is $B_{1g}$, 4 is $B_{1u}$, 5 is $B_{2g}$,
6 is $B_{2u}$, 7 is $B_{3g}$ and 8 is $B_{3u}$);

$n_1$-$n_{12}$: \trove\ vibrational quantum numbers (QN); see Table~\protect\ref{tab.vib_comp} for the correlation with the normal QNs;

$\Gamma_{\text{vib}}$: Symmetry of vibrational component of state in D$_{2h}$(M);

$K$: Projection of $J$ on molecule-fixed $z$-axis;

$\tau_{\text{rot}}$: Rotational parity (0 or 1);

$\Gamma_{\text{rot}}$: Symmetry of rotational component of state in D$_{2h}$(M).

\end{table}

{\renewcommand{\arraystretch}{1.1}
\renewcommand{\tabcolsep}{0.5cm}
\begin{table}
\caption{Extract from .trans file for MaYTY line list.}
\begin{tabular}{c c c}
\hline
$f$ & $i$  & $A_{fi}$ \\
\hline
      &       &             \\
49589 & 44178 &  2.4146E-07 \\
49590 & 44178 &  2.1037E-05 \\
12140 & 44178 &  2.1033E-05 \\
49591 & 44178 &  1.8719E-05 \\
\hline
\end{tabular}
\label{tab.tran}

$f$: Upper state ID;

$i$: Lower state ID;

$A_{fi}$: Einstein A coefficient (in s$^{-1}$).
\end{table}

\subsection{Validation}

The MaYTY line list contains nearly 50 billion (49~841~085~051) transitions between over 45 million (45~446~267) states.

Fig. \ref{fig:compare_whole} and \ref{fig:compare_HITRAN} compare
the MaYTY line list to empirical intensities from the HITRAN database
at 296 K.  To take into account the abundance of the
$^{12}$C$_2$$^1$H$_4$ isotopologue we divide the HITRAN intensities by
0.97729 to obtain the value for pure $^{12}$C$_2$$^1$H$_4$.
Fig.~\ref{fig:compare_whole} gives an overview of our line list
compared to HITRAN. HITRAN is currently missing data for the 1600 --
2750 cm$^{-1}$ region which contains the relatively intense $\nu_7 +
\nu_8$ band amongst others. For this region  Fig.
\ref{fig:compare_HITRAN} compares our line list to that of
\citet{16ReDeNi} from the TheoReTS database. HITRAN also currently
does not have any data above 3500 cm$^{-1}$.

\begin{figure}
\centering
\includegraphics[scale=0.6]{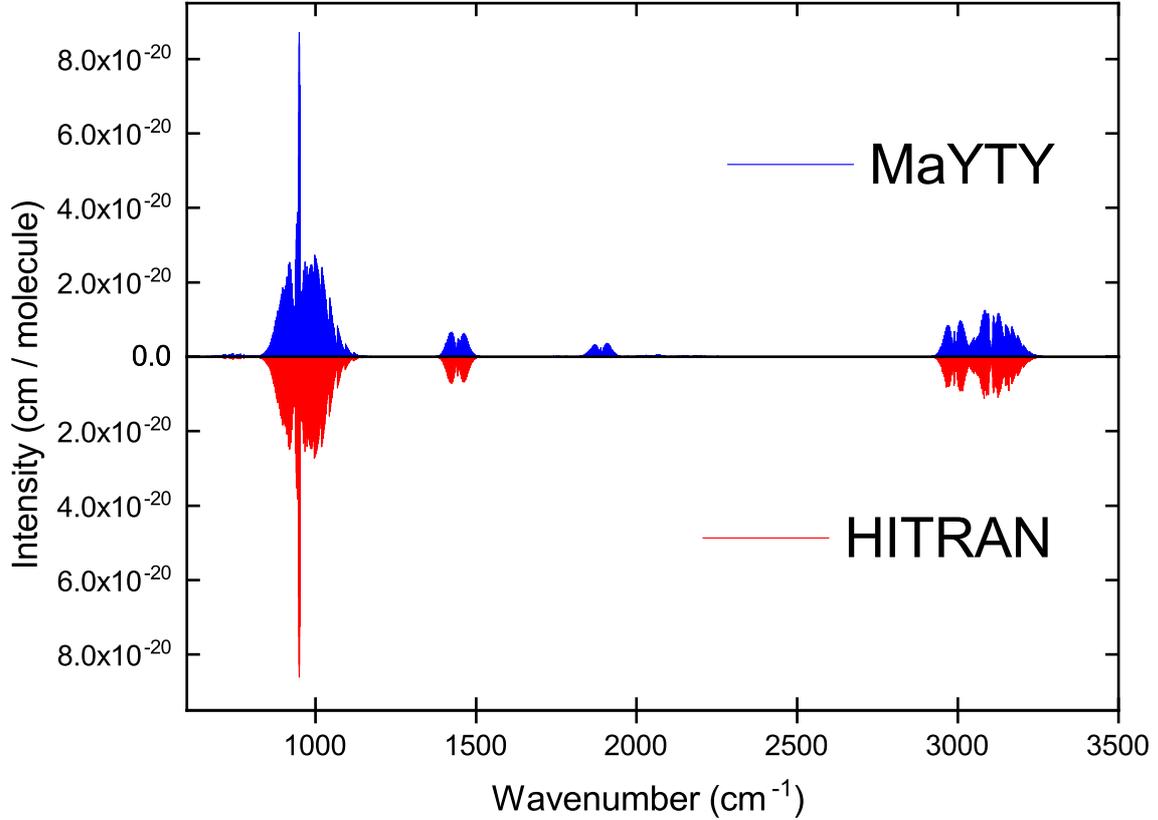}
\caption{Overview of absolute intensities of MaYTY compared to HITRAN data at 296~K. }
\label{fig:compare_whole}
\end{figure}

\begin{figure}
\centering
  \begin{tabular}{@{}cc@{}}
    \includegraphics[width=.5\textwidth]{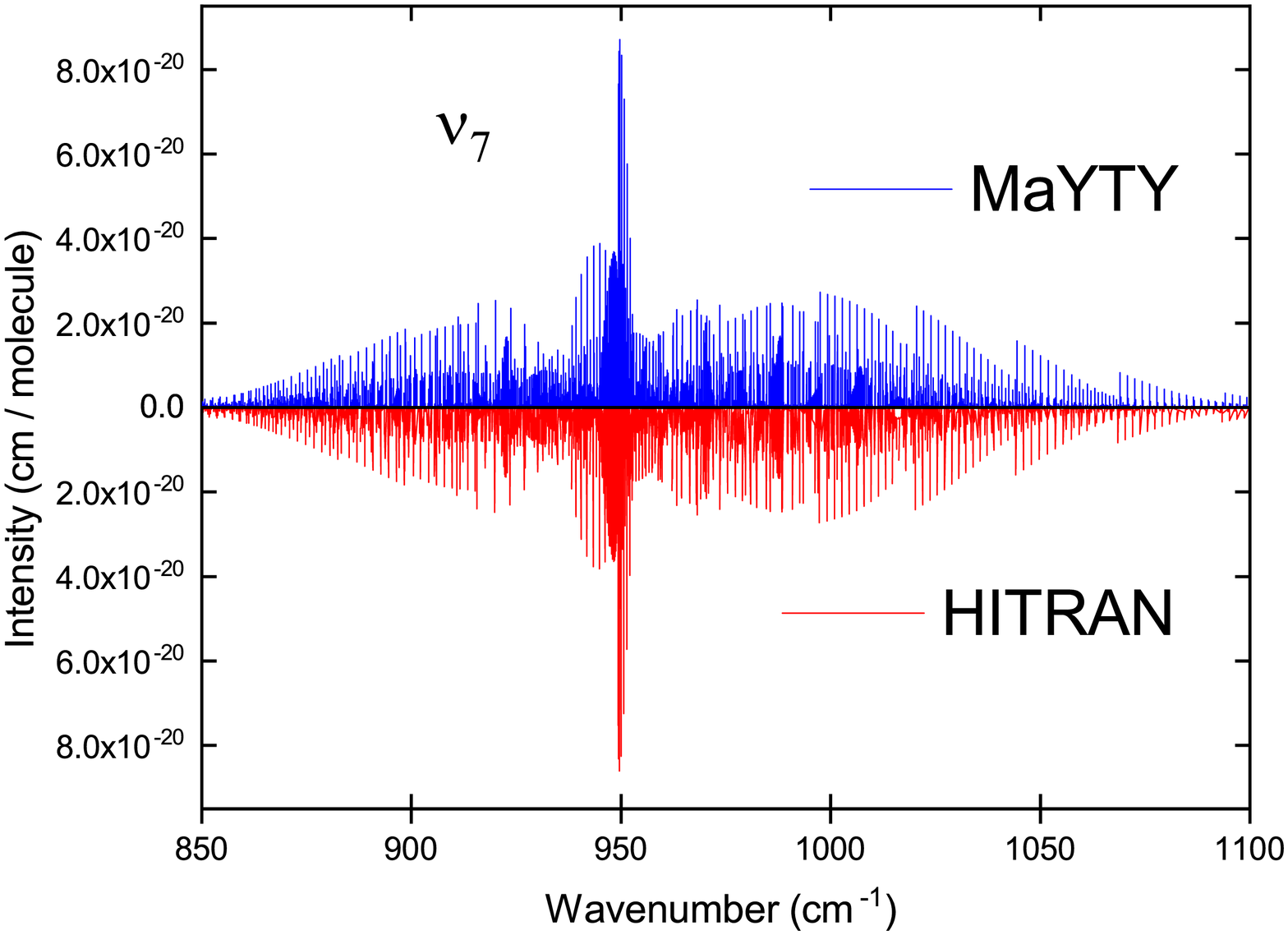} &
    \includegraphics[width=.5\textwidth]{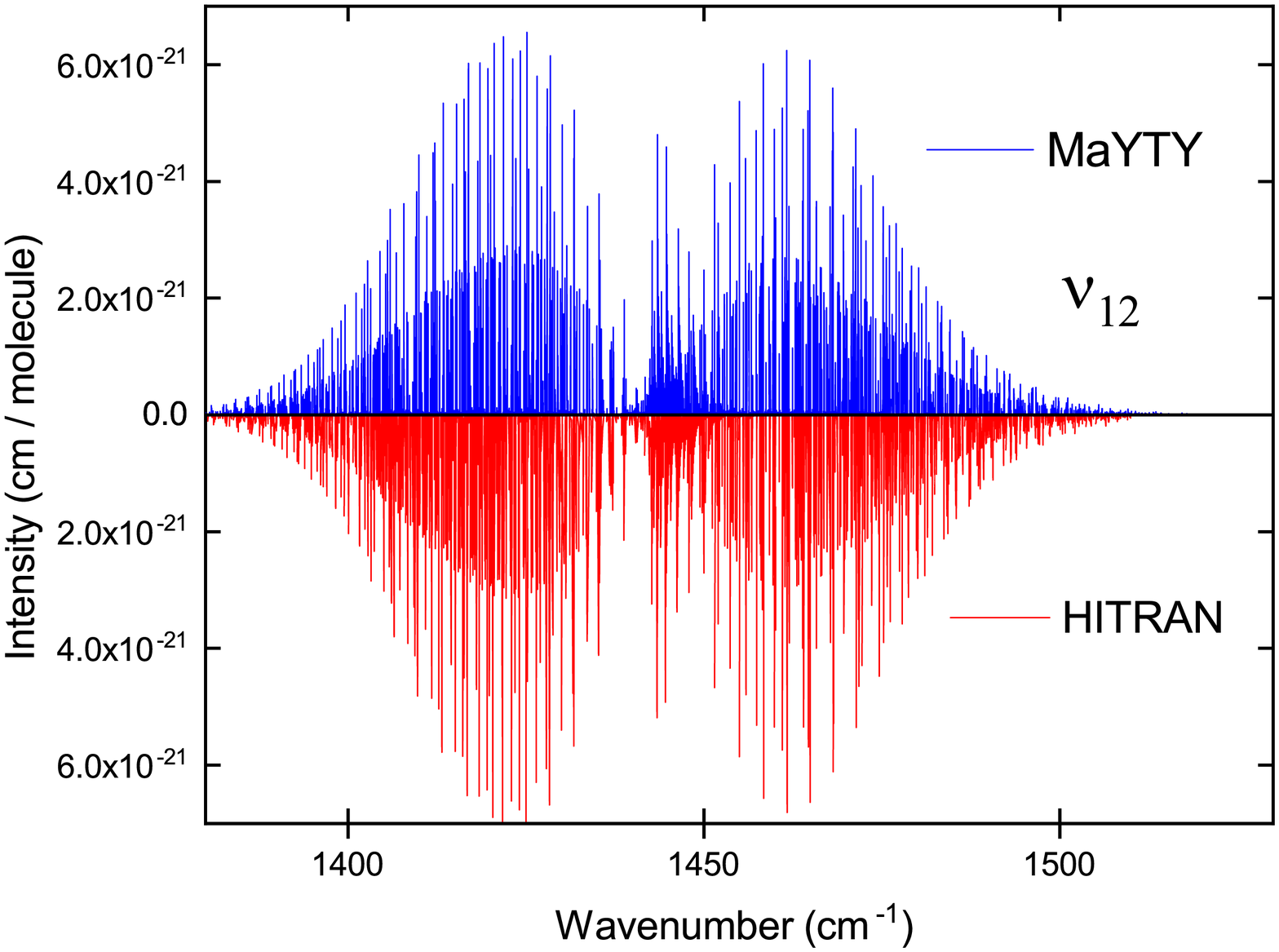} \\
    \includegraphics[width=.5\textwidth]{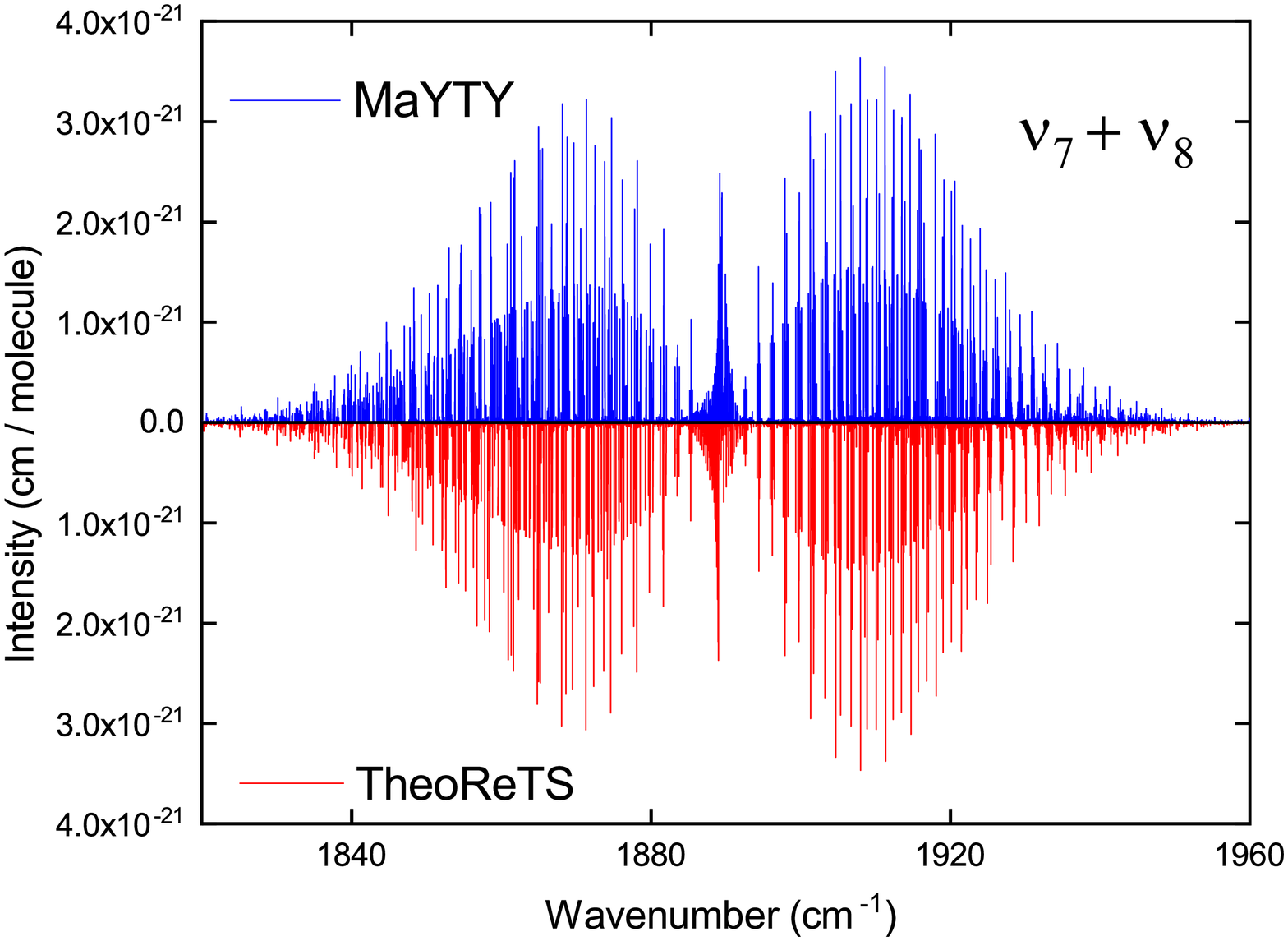} &
    \includegraphics[width=.5\textwidth]{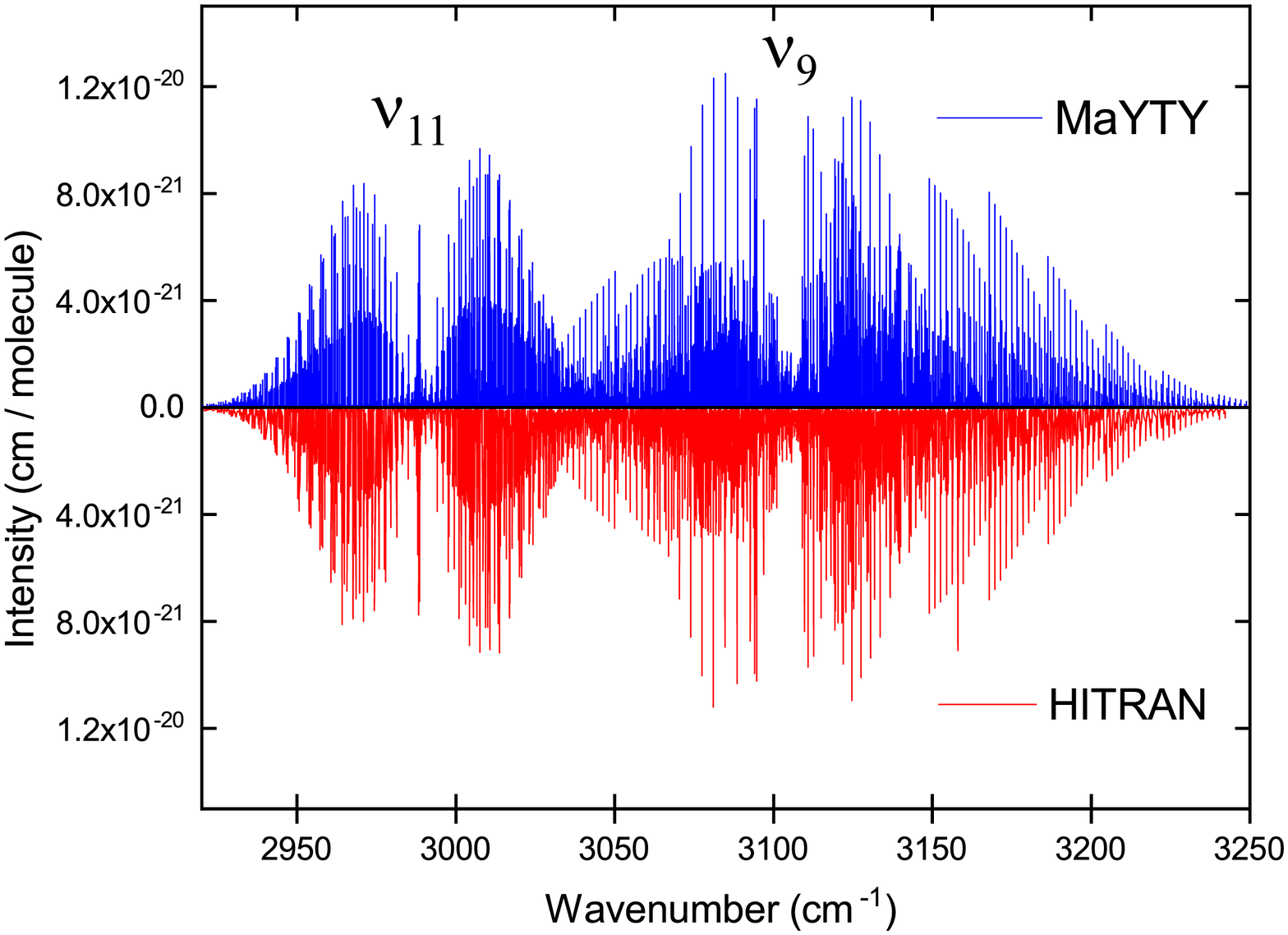} \\
  \end{tabular}
  \caption{Detailed comparison of the MaYTY line list those from HITRAN and TheoReTS at 296~K. The strongest bands are labelled
  in each plot.}
  \label{fig:compare_HITRAN}
\end{figure}

As a further comparison of the MaYTY line list to experiment we compare
cross sections to those in  the PNNL database \citep{04ShJoSa} in Fig. \ref{fig:compare_PNNL}.
The PNNL spectrum is a composite recorded at 298 K and re-normalized
for 296 K. The ethylene used was 99.5 \% pure. For comparison with our line list PNNL cross sections were multiplied by
$9.28697 \times 10^{-16}$/0.97729 to convert to cm$^2$ molecule$^{-1}$ units and account for the $^{12}$C$_2$$^1$H$_4$
isotopologue. We simulated the spectrum using a resolution of 0.1 cm$^{-1}$ using a Voigt profile with a half-width
half-maximum (HWHM) value of 0.1 cm$^{-1}$. The PNNL spectrum allows a comparison to our calculated line list up to around
6200 cm$^{-1}$.

\begin{figure}
\centering
  \begin{tabular}{cc}
    \includegraphics[width=.5\textwidth]{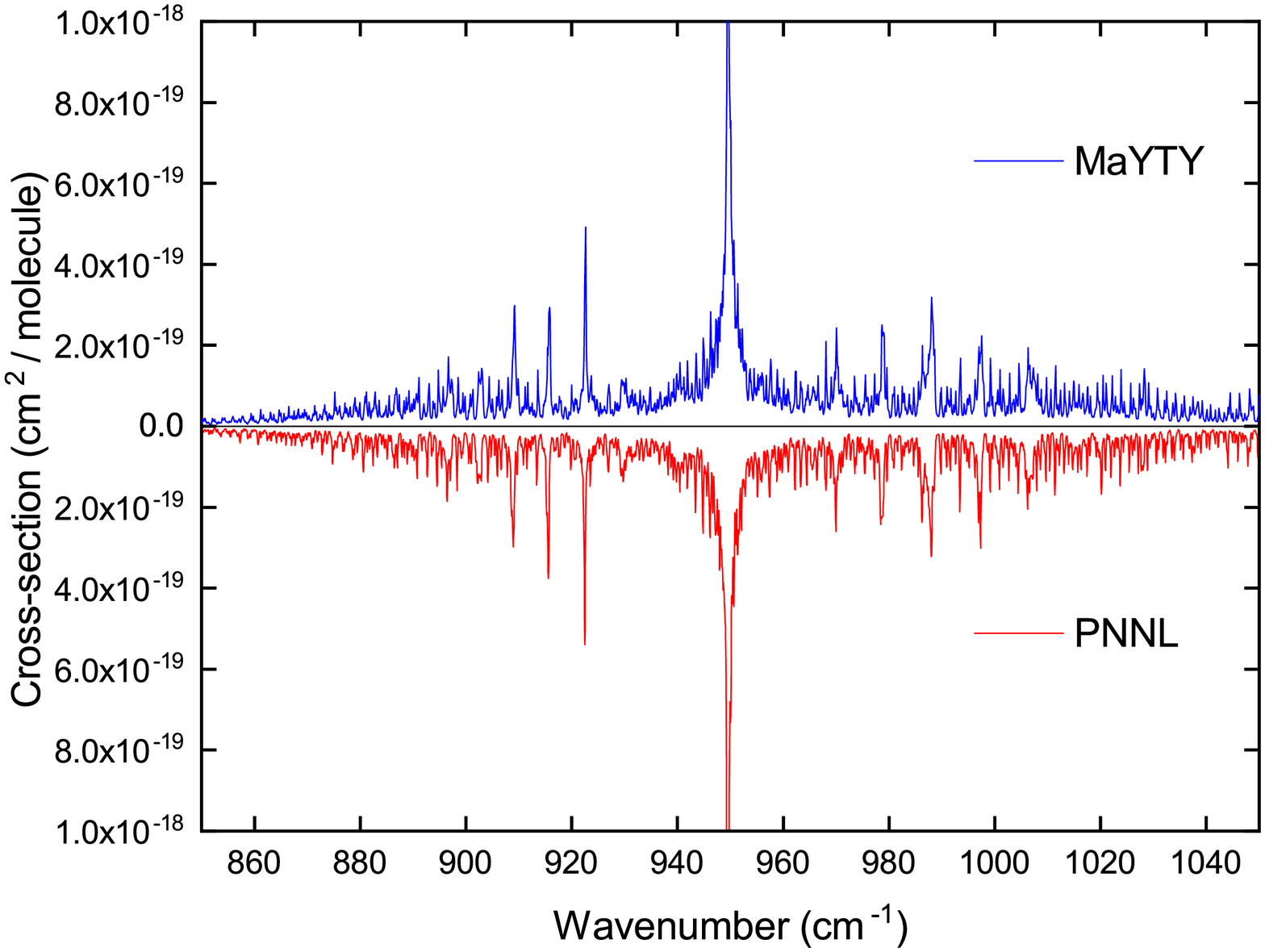} &
    \includegraphics[width=.5\textwidth]{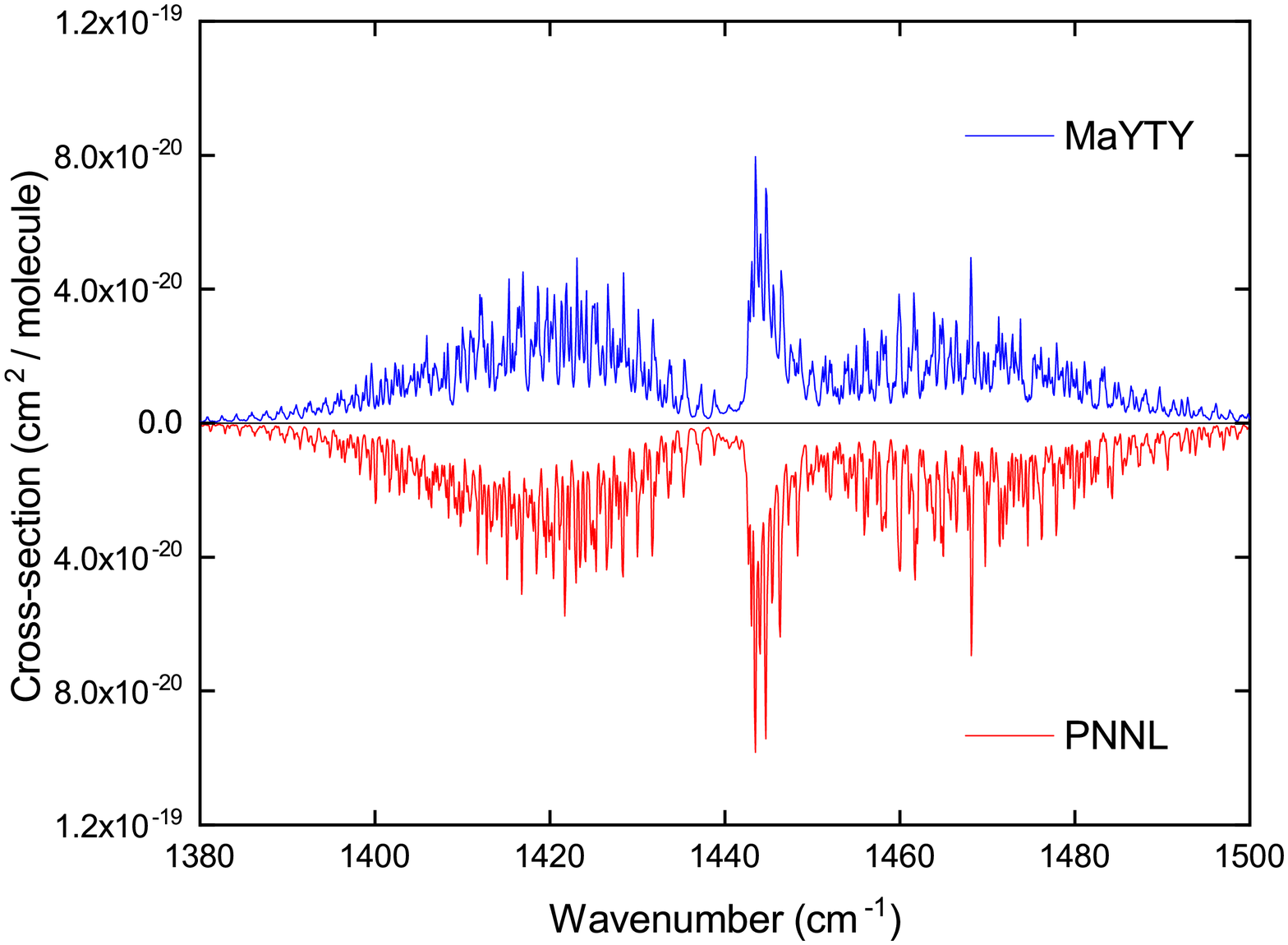} \\
    \includegraphics[width=.5\textwidth]{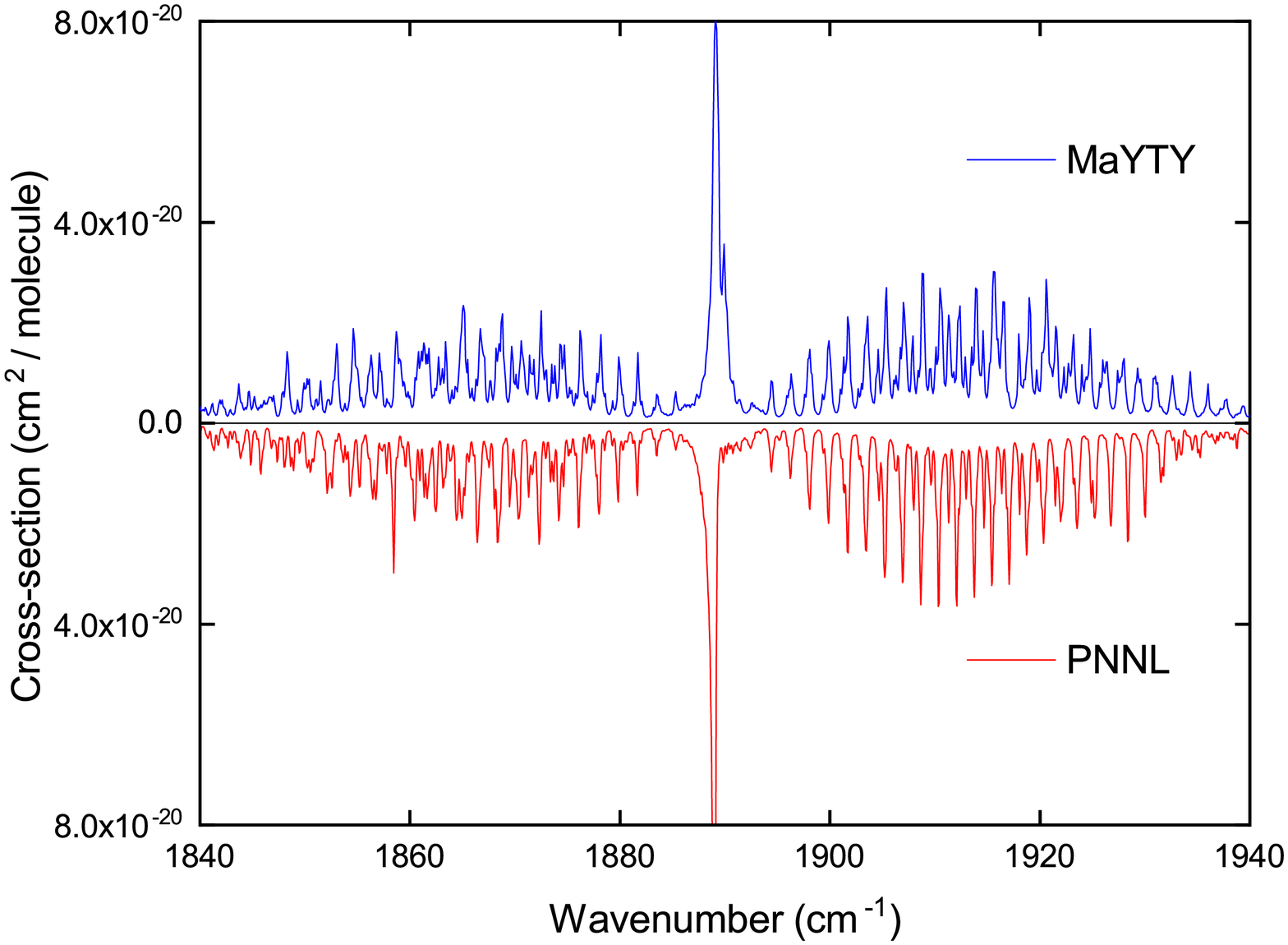} &
    \includegraphics[width=.5\textwidth]{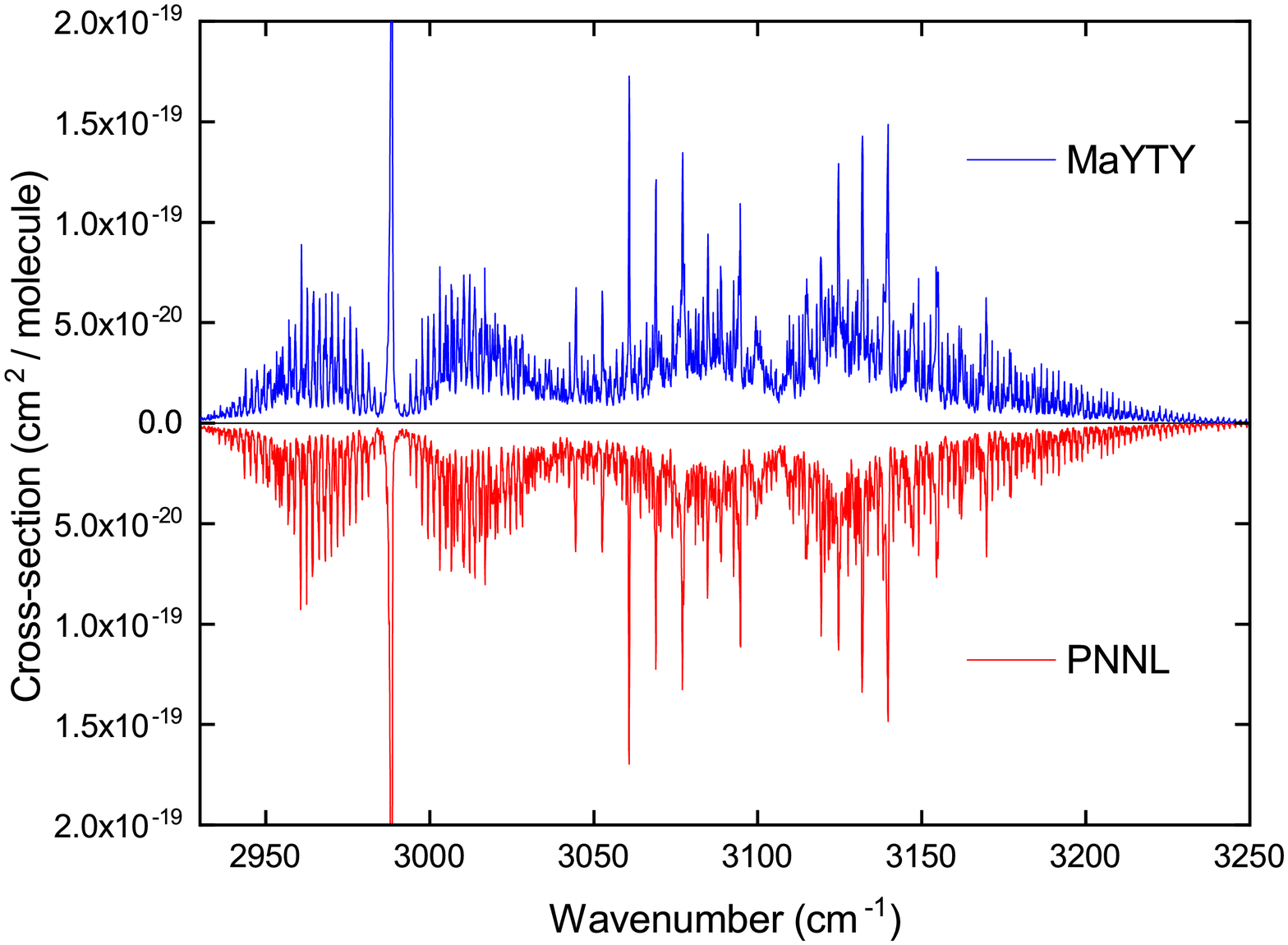} \\
    \includegraphics[width=.5\textwidth]{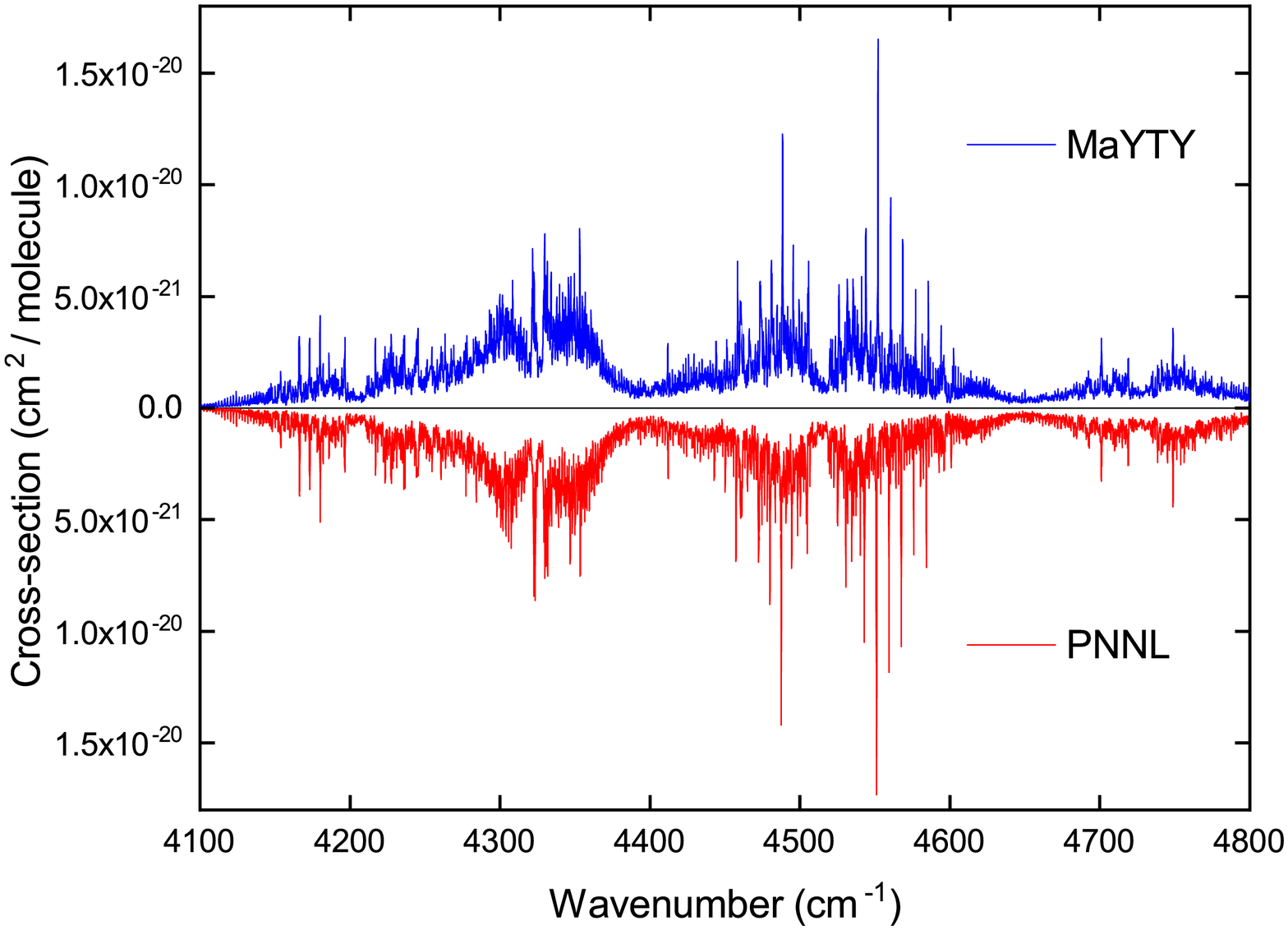} &
    \includegraphics[width=.5\textwidth]{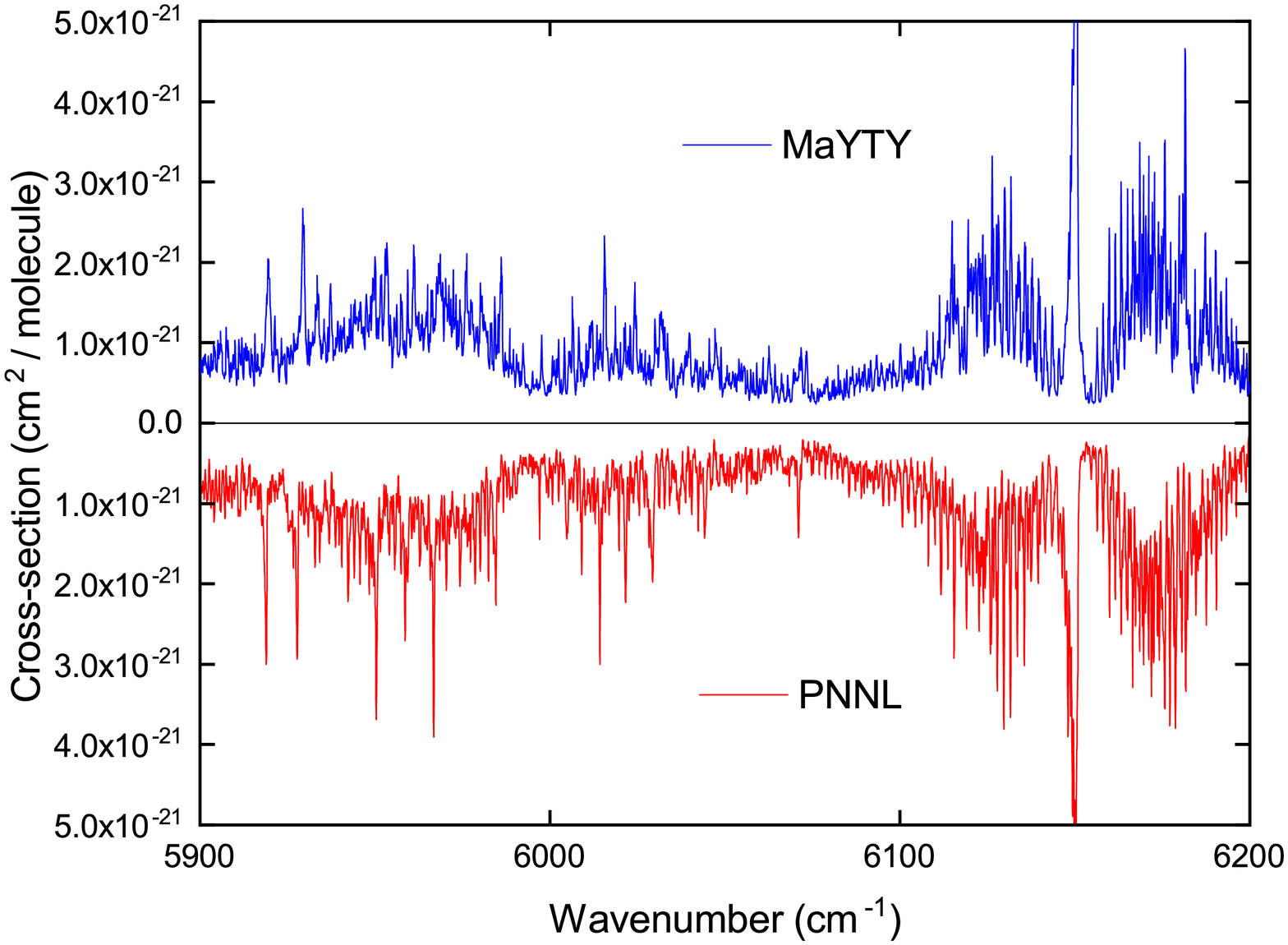} \\
  \end{tabular}
  \caption{Comparisons of MaYTY line list to PNNL cross sections at 296~K}
  \label{fig:compare_PNNL}
\end{figure}

Fig. \ref{fig:temp_dep} shows the temperature dependence of the absorption cross sections for the MaYTY line list
simulated using a resolution of 5 cm$^{-1}$ where again a Voigt profile with a HWHM of 0.1 cm$^{-1}$ was used.
While regions of weak absorption at 296 K increase by an order of magnitude or more as the temperature is increased,
the overall band structure of the absorption does not change greatly with temperature. This behaviour contrasts
with other molecules, such as methane \citep{jt564}, whose band shapes show a strong temperature dependence.

\begin{figure}
\centering
\includegraphics[scale=0.6]{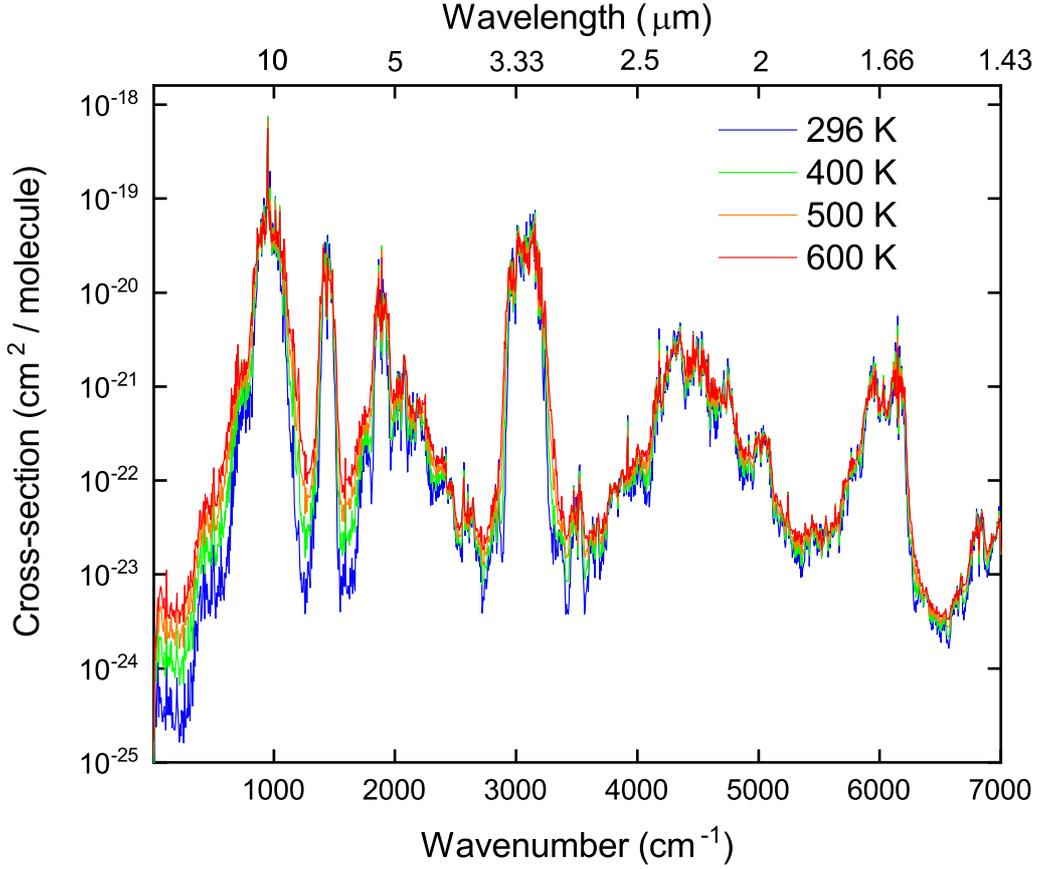}
\caption{Infrared absorption spectrum of $^{12}$C$_2$$^1$H$_4$:  Temperature dependence of MaYTY line list.}
\label{fig:temp_dep}
\end{figure}

\section{Vibrational cross sections: realistic band shapes}
\label{sec:top-up}

Our final ro-vibrational line list for C$_2$H$_4$ is limited to the frequency range of 0 -- 7~000~\cm\
and the temperatures up to about 700~K. On the other hand, our vibrational line list, which we used to assist the basis set pruning, has much larger coverage, both in terms of frequency and temperature ranges. Here we show how to use this more complete vibrational line list to (i) top-up the ro-vibrational line list
with missing opacities and (ii) directly simulate realistic spectra
at high temperatures for large polyatomic molecules. 

The vibrational line list consists of the vibrational Einstein coefficients $A_{fi}^{(J=0)}$ generated using Eq.~\eqref{eq.vib-einsteinA}, which are stored using the ExoMol line list format: the vibrational .trans  has the same format as in Table~\ref{tab.vib-tran}, i.e. with the upper state ID, lower state ID and the third column containing $A_{fi}^{(J=0)}$. The vibrational .states file contains all the vibrational energy term values labelled with their IDs, as in Table~\ref{tab.states}; the only difference with the ro-vibrational .states file is that the statistical weights are all set to 1 according with Eqs.~\eqref{eq.vibint} and \eqref{eq.vib-einsteinA}. The vibrational line list in this format can be used together with {\sc ExoCross} to generate absorption vibrational intensities using Eq.~\eqref{eq.vibint} (instead of its ro-vibrational analogy in Eq.~\eqref{eq.intensity}). Another potential application of our extensive vibrational line list for ethylene is to generate spectra using the spectroscopic tool \textsc{PGOPHER} \citep{PGOPHER}. One of the recent features of \textsc{PGOPHER} is to import band centers and and transition moments from an external  vibrational line list.

Here we use the hot vibrational line list for C$_2$H$_4$ to produce temperature-dependent vibrational cross sections by `broadening' the corresponding band intensity with suitable band profiles. The vibrational cross sections should, at least approximately, conserve the opacity stored in each vibrational band and thus offer an approximate but simple way of simulating molecular opacity.

In fact, it is common in applications involving large polyatomic molecules to use  vibrational intensities for modelling molecular absorption, where Lorentzian or Gaussian functions are used as band profiles. There are also more realistic but elaborate alternatives to represent the band profiles, such as, for example the narrow band approach \citep{15CoLixx}. Here we develop a three-band model, where different vibrational bands (perpendicular and parallel) are modelled using three realistic basic shapes, corresponding to three components of the vibrational dipole moment $\bar{\mu}_{\alpha}$ of ethylene.

{\renewcommand{\arraystretch}{1.1}
\renewcommand{\tabcolsep}{0.5cm}
\begin{table}
\caption{Extract from vibrational .trans file.}
\begin{tabular}{rrr}
\hline
$f$ & $i$  & $\mu_{fi}$ \\
\hline
      &       &             \\
       65040     &      1 &   5.44101247E-07 \\
        1040     &      1 &   3.33556008E-16 \\
       85543     &      1 &   1.81580111E-16 \\
      127077     &      1 &   7.88402351E-16 \\
       63744     &      1 &   1.92578661E-07 \\
        8779     &      1 &   1.45418720E-16 \\
       43097     &      1 &   1.23733815E-20 \\
\hline
\end{tabular}
\label{tab.vib-tran}
\mbox{}\\
$f$: Upper state ID;\\
$i$: Lower state ID; \\
$\mu_{fi}$: Vibrational transition moment (in D).
\end{table}


As indicated in Table~\ref{tab.vib_comp}, only the $\nu_7$, $\nu_9$,
$\nu_{10}$, $\nu_{11}$, $\nu_{12}$ bands are IR active: $\nu_{9}$ and
$\nu_{10}$ (see Figs.~\ref{fig:compare_HITRAN}) are parallel bands as
they possess the same symmetry $B_{1u}$ as the $z$ component of the
molecular dipole moment $\boldsymbol\mu$. The perpendicular bands
$\nu_{11}$ and $\nu_{12}$ are of the type $B_{2u}$ (corresponding to
$\bar{\mu}_y$), while the perpendicular band $\nu_7$ is of the type
$B_{3u}$ ($\mu_x$). These three band types ($B_{1u}$, $B_{2u}$ and
$B_{3u}$) have different shapes, which we use as templates to model
all other vibrational bands of C$_2$H$_4$. We select the three
strongest fundamental bands, one for each type: $\nu_{12}$ ($B_{1u}$),
$\nu_{9}$ ($B_{2u}$) and $\nu_{7}$ ($B_{3u}$), and use the
corresponding ro-vibrational cross sections at different temperatures
to construct three temperature-dependent, normalised band profiles as
follows.  For each temperature and band in question the corresponding
cross-sections on a grid of 1~\cm\ are normalised and shifted to have
the center at $\tilde{\nu}=0$. Three profile templates for $T=500$~K
and $T=1500~K$ are shown in Fig.~\ref{fig:templates}.  The $T=500$~K
profiles were generated using the MaYTY line list in conjunction with
the Voigt line profile with HWHM=0.1 cm$^{-1}$. For the $1500$~K
temperature case our line list is rotationally incomplete ($J_{max} =
78$), therefore we used the effective Hamiltonian approach to generate
the corresponding band profiles with significantly higher $J_{\rm max}
= 120$. Towards this we employed {\sc PGOPHER}  together with
the $\nu_{7}$, $\nu_{9}$ and $\nu_{12}$ spectroscopic constants from
\citet{98BaGeHe.C2H4} and \citet{98RuFiKh.C2H4}, and a Voigt line profile with
HWHM=0.1 cm$^{-1}$.

These profiles   are then applied for the vibrational cross sections at the temperature in question by using the symmetry multiplication rule: if $\Gamma_{i}$ and  $\Gamma_f$ are, respectively, the symmetries  of the initial and upper states and $\Gamma_\alpha$  is the symmetry of the dipole moment component $\bar{\mu}_{\alpha}$, for an IR active band $\tilde{\nu}_{fi}^{(J=0)}$ the following relation holds  \citep{98BuJexx}:
$$
\Gamma_{\alpha} = \Gamma_{i} \otimes \Gamma_{f}.
$$
Note that the equal sign here (not $\in$) is due to D$_{2h}$(M) being an Abelian symmetry group.
We thus use this rule to choose between the $B_{1u}$, $B_{2u}$ or $B_{3u}$  templates when generating cross sections for specific bands $\tilde\nu_{fi}^{(J=0)}$. This rule, however, does not always hold: a large number of forbidden (and weak) bands have  non-zero intensities due to interactions between vibrational states. In such cases we use a simple  Lorentzian band profile with HWHM of 60~\cm.

\begin{figure}
\centering
\includegraphics[scale=0.28]{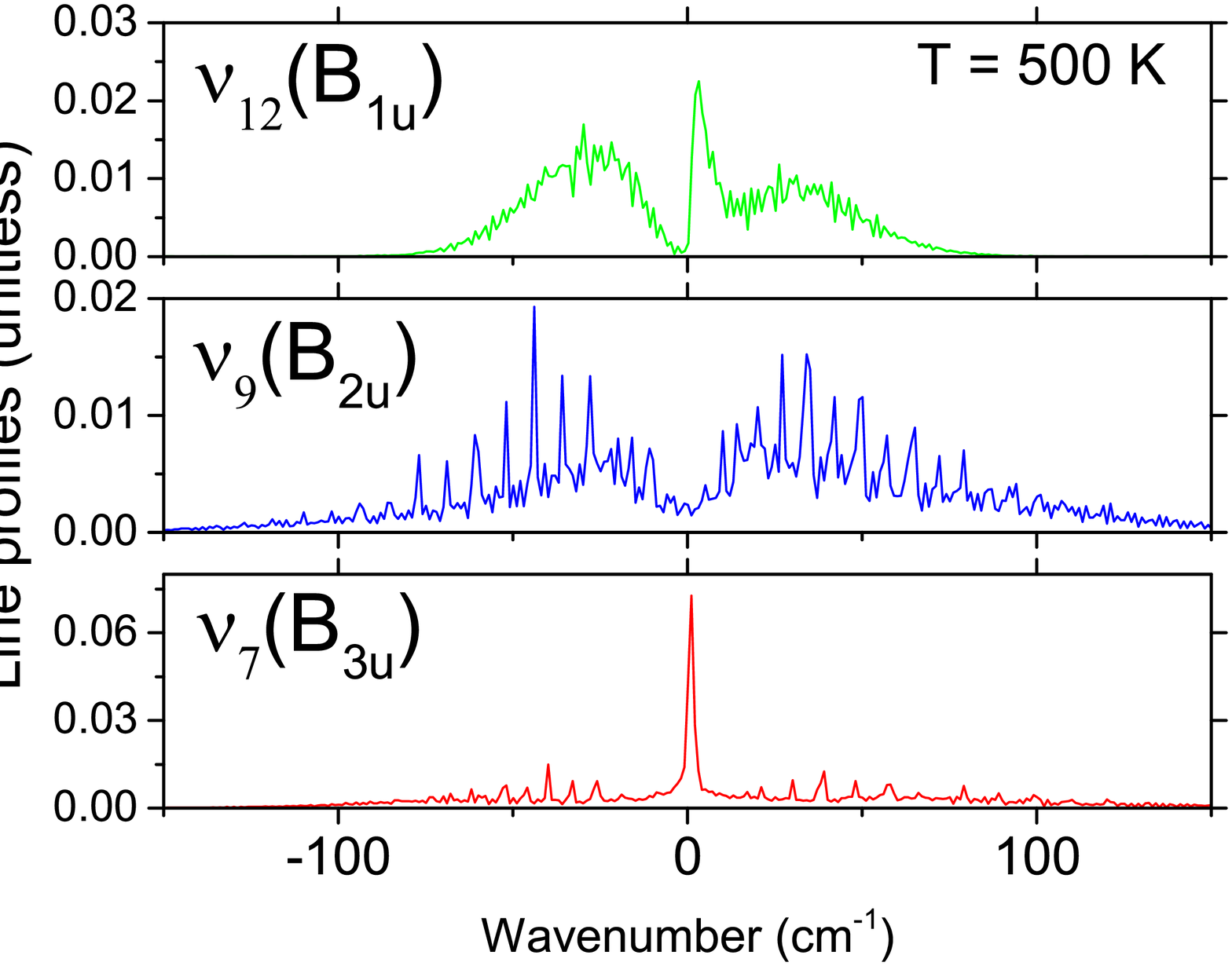}
\includegraphics[scale=0.28]{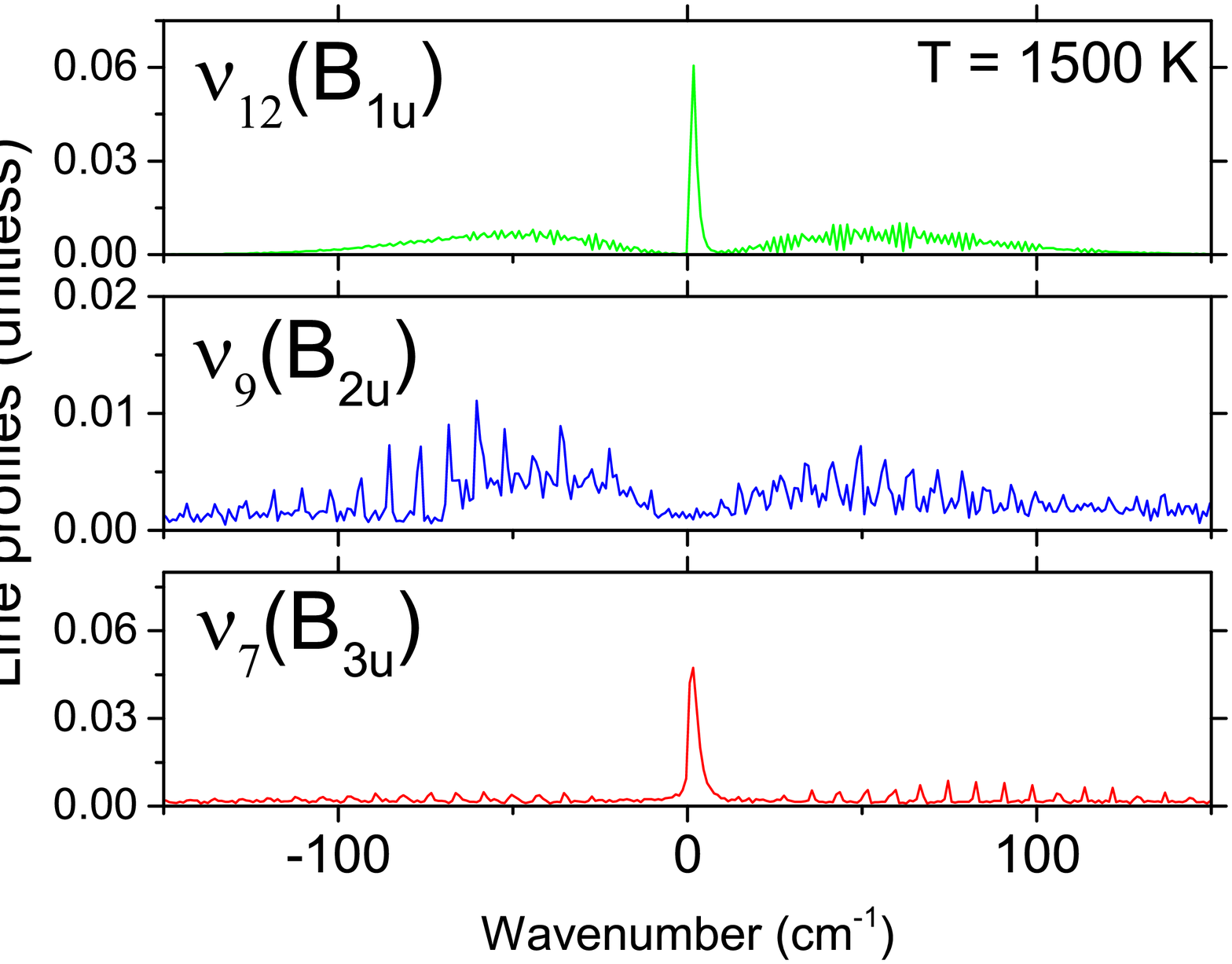}
\caption{Normalized $\nu_{12}$,  $\nu_{9}$ and $\nu_{7}$ band profiles with a HWHM=0.1~\cm. Left:   the $T=500$~K profiles were generated using the MaYTY line list with the Voigt profile. Right: The $T=1500$~K profile was generated using {\sc PGOPHER} with the Lorentzian line broadening and spectroscopic constants from \protect\citet{98BaGeHe.C2H4} and \citet{98RuFiKh.C2H4}.}
\label{fig:templates}
\end{figure}

An example of vibrational cross sections of C$_{2}$H$_4$
at $T=500$~K  and $T=1500$~K  generated using this methodology is  shown in Fig.~\ref{fig:T=500K}, where they are also compared to the ro-vibrational
cross sections. The vibrational cross sections are more complete and also provide larger coverage (here shown up to 10~000~\cm). For example, the lower display on Fig.~\ref{fig:T=500K} shows the predicted opacity of  C$_2$H$_4$ at $T=1500$~K by using our vibrational cross section technique compared to the MaYTY intensities, which are incomplete at $T=1500$~K.

The methodology of combining realistic band profiles with vibrational intensities can be especially
useful for larger polyatomic molecules, where the size of the calculations becomes prohibitive. This requires knowing the ro-vibrational spectra of the three fundamental bands to generate the realistic band profiles, for which we took advantage of having the complete, ro-vibrational line list. In practical applications when this is not accessible,  these profiles could be modelled using effective rotational methods, using for example {\sc PGOPHER} \citep{PGOPHER} as we demonstrated , which only requires the corresponding spectroscopic constants of these (up to) three fundamental bands.

\begin{figure}
\centering
\includegraphics[scale=0.45]{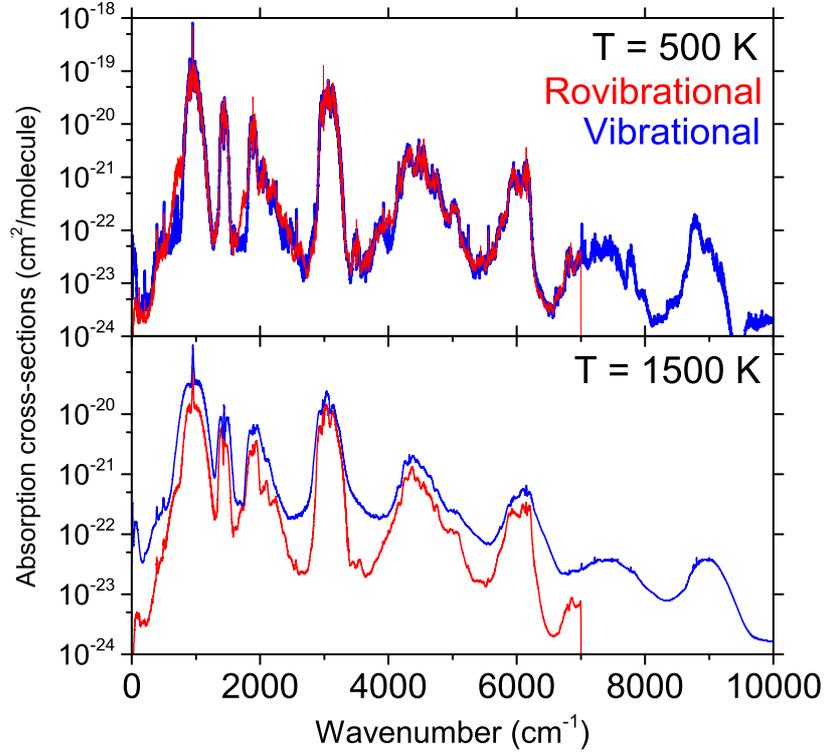}
\caption{Comparing the ro-vibrational and vibrational intensities of  C$_2$H$_4$ at $T=500$~K and $T=1500$~K: The `ro-vibrational' cross sections of C$_2$H$_4$ were computed using the MaYTY line list and the Gaussian line profile with HWHM=1~\cm.  The `vibrational' cross sections were computed using the vibrational line list with the normalized band profiles scheme (see text). }
\label{fig:T=500K}
\end{figure}

The temperature dependent vibrational cross sections can  be useful for evaluating opacities of molecules (especially at higher temperatures) when completeness is more important than high accuracy.  The approximations used for vibrational intensities are: (i) the rotational and vibrational
degrees of freedom are independent and (ii) lower
resolution is assumed.

Due to the missing interaction between the rotational and vibrational degrees of freedom, this vibrational methodology is not capable of reconstructing some  forbidden bands, which are caused by this interaction. This is evident in Fig.~\ref{fig:T=500K}, where some weaker parts are missing.  It is important to note that the vibrational band
intensity of a given vibrational band computed using
Eq.~\eqref{eq.vibint} is the same as the corresponding integrated
ro-vibrational intensities from Eq.~\eqref{eq.intensity}, at least if the
interaction with other vibrational bands is ignored. Thus although the
vibrational intensity treatment is highly approximate, it should be better for preserving the opacity in simulations.

In line with our `hybrid'-methodology \citep{jt698}, the generated vibrational cross sections can be now divided
into the strong and weak parts, with the latter representing the
absorption, missing from our line list due to the vibrational basis
set pruning. These `weak' vibrational bands form absorption `continuum' cross sections and can be used to compensate for missing absorption when higher temperatures or larger spectroscopic coverage is required.
Fig.~\ref{fig:vib-super-lines} shows this absorption continuum of C$_2$H$_4$ at $T=500$~K up to 16~000~\cm\ as well as the the ro-vibrational cross sections  generated using the MaYTY line list below 7~000~\cm.

\begin{figure}
\centering
\includegraphics[scale=0.5]{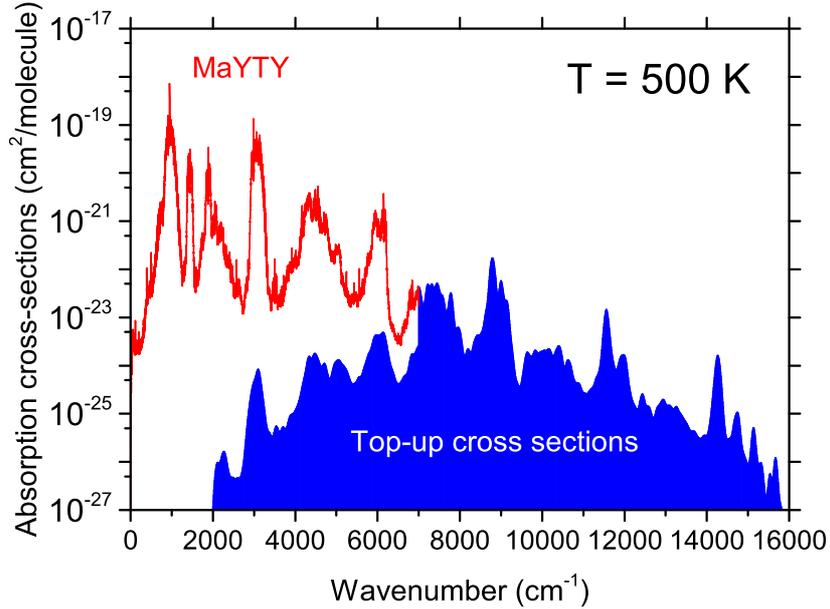}
\caption{The vibrational `top-up' cross sections of C$_2$H$_4$ computed using the vibrational line list  with the 3-band model at $T=500$~K (dark blue area) and the ro-vibrational intensities generated using the ExoMol line list with the Gaussian line profile of HWHM=1~\cm\ (red line). }
\label{fig:vib-super-lines}
\end{figure}


\section{Conclusions}
\label{sec:conclusions}

We have produced a new line list for $^{12}$C$_2$$^1$H$_4$ called MaYTY. Energy levels were calculated variationally using the
\trove\ program on a refined potential energy surface and transition intensities calculated with a new
\textit{ab initio} dipole moment surface. 
The MaYTY line list includes transitions between ro-vibrational states with $J\leq78$ and covers the frequency region up to 7000 cm$^{-1}$. Based on analysis of the partition function the variational line list is applicable up to 700 K beyond which opacity will be
underestimated. Our line list is in good agreement with experimental intensities from the HITRAN database and
experimental cross sections from the PNNL database. The
MaYTY line list is available
from the CDS (http://cdsarc.u-strasbg.fr) and ExoMol (www.exomol.com)
data bases.

To extend the temperature range of applicability of the line list we have implemented a new approximate method of producing
absorption cross sections from vibrational ($J=0$)  energies and transition moments. The vibrational line list of C$_2$H$_4$ is also provided as part of the  MaYTY data set. Using this approach we have also generated vibrational cross sections of C$_2$H$_4$ covering the wavenumber range up to 12~000~\cm\ and covering temperatures up to $T=1500$ K.
The vibrational cross sections based on the 3-band model  can be also generated using a vibrational version of {\sc ExoCross}. This program, vibrational cross sections and vibrational line list for ethylene are available from the CDS and ExoMol data bases. The $J=0$ method proposed should be useful for calculating  opacities of larger molecules where high $J$ variational calculations are extremely challenging.

\section{Acknowledgements}

This work was supported by the UK Science and Technology Research Council (STFC) No. ST/M001334/1 and the COST action MOLIM No. CM1405.  This work made extensive use of UCL's Legion and DARWIN and COSMOS high performance computing facilities provided by DiRAC for particle physics, astrophysics and cosmology and supported by STFC and BIS.

\label{lastpage}

\bibliographystyle{mnras}


\end{document}